\documentclass{article}

\usepackage{PRIMEarxiv}

\usepackage[utf8]{inputenc} 
\usepackage[T1]{fontenc}    
\usepackage{hyperref}       
\usepackage{url}            
\usepackage{booktabs}       
\usepackage{nicefrac}       
\usepackage{microtype}      
\usepackage{lipsum}
\usepackage{fancyhdr}       
\usepackage{graphicx}       
\graphicspath{{media/}}     

\usepackage{siunitx}
\usepackage{dsfont}
\usepackage{subcaption}
\usepackage{lineno}
\usepackage{bm}
\usepackage[super,sort&compress]{natbib}
\usepackage{authblk}

\makeatletter
\renewcommand\NAT@open{[}
\renewcommand\NAT@close{]}
\renewcommand\@cite[1]{\textsuperscript{[#1]}}
\makeatother

\usepackage{amssymb}
\usepackage{comment}
\usepackage{soul}
\usepackage{multirow}%
\usepackage{amsmath,amssymb,amsfonts}%
\usepackage{amsthm}%
\usepackage{mathrsfs}%
\usepackage[title]{appendix}%
\usepackage{xcolor}%
\usepackage{textcomp}%
\usepackage{manyfoot}%
\usepackage{algorithm}%
\usepackage{algorithmicx}%
\usepackage{algpseudocode}%
\usepackage{listings}%

\theoremstyle{thmstyleone}%
\theoremstyle{thmstyletwo}%

\theoremstyle{thmstylethree}%

\usepackage{caption}

\renewcommand{\figurename}{Fig.}
\makeatletter
\renewcommand{\keywords}[1]{%
  \par\noindent\textbf{Keywords:} #1
}
\makeatother

\title{Radiosonde-constrained reconstructions reveal a weakening Northern Hadley circulation}

\author[ ]{%
  \begin{minipage}{0.4\textwidth}
    \centering
    \textbf{Matic Pikovnik}\textsuperscript{1} \\
    \texttt{matic.pikovnik@fmf.uni-lj.si}
  \end{minipage}%
  \vspace{3mm}
}
\author[ ]{%
  \begin{minipage}{0.4\textwidth}
    \centering
    \textbf{\v{Z}iga Zaplotnik}\textsuperscript{2,1} \\
    \texttt{ziga.zaplotnik@ecmwf.int}
  \end{minipage}%
  \vspace{3mm}
}

\affil[1]{Faculty of Mathematics and Physics, University of Ljubljana, \protect\\Jadranska 19, 1000 Ljubljana, Slovenia\vspace{4mm}}
\affil[2]{European Centre for Medium-range Weather Forecasts, \protect\\Robert-Schuman-Platz 3, 53175 Bonn, Germany}

\date{} 

\begin{document}
\maketitle

\begin{abstract}
The Northern Hadley cell (NHC) is a fundamental component of Earth's atmospheric circulation, governing precipitation patterns affecting nearly four billion people. Despite its importance, the sign of recent multidecadal trends in NHC strength remains unresolved. Climate models consistently simulate a weakening, whereas reanalyses have suggested an opposing strengthening. Here, we constrain this discrepancy using the global radiosonde record. To assess the NHC, we reconstruct the three-dimensional meridional wind from sparse radiosonde observations using a masked autoencoder graph neural network and apply an identical reconstruction to five modern reanalyses, sampled at the same locations. This paired reconstruction framework reveals a systematic underestimation of climatological NHC strength across all reanalyses, corroborated in ERA5 by systematic data assimilation increments that persistently strengthen the circulation. Most importantly, our radiosonde-based reconstructions provide vertically resolved observational evidence of a statistically significant weakening of the NHC since 1980, reconciling observations with climate model projections. The weakening is consistently reproduced by all reanalysis-based reconstructions and is robust across training datasets and analysis periods, strengthening confidence in projected changes in the Hadley circulation. More broadly, this study establishes a temporally homogeneous reconstruction framework for evaluating large-scale circulation changes and assessing both reanalysis products and climate model projections.
\end{abstract}

\keywords{Hadley circulation strength, radiosonde observations, reanalyses, masked autoencoder, graph neural network, bias, trends, climate change, data assimilation}

\section{Introduction}\label{sec1}

The Hadley circulation is a fundamental component of Earth's atmospheric system, characterised by a large‑scale meridional overturning with rising motion near the Intertropical Convergence Zone (ITCZ) and subsidence in the subtropics. It consists of two Hadley cells that transport moisture, heat, and momentum between the ITCZ and higher latitudes\textsuperscript{\cite{Peixoto1992PhysicsClimate}}. Variations in their strength directly influence regional hydroclimates\textsuperscript{\cite{Lionello2024}}, affecting precipitation patterns and water availability across densely populated regions of the globe\textsuperscript{\cite{Held2006,Laua2015,ipcc_ar6_wg1_ch8}}. 
Understanding the historical mean state and evolution of the Hadley circulation is therefore essential for constraining its response to anthropogenic forcing and for establishing confidence in climate model projections used to guide climate-change adaptation.

Despite its importance, the sign of recent multidecadal trends in Northern Hadley cell (NHC) strength remains uncertain\textsuperscript{\cite{Lionello2024,ipccar6wg1}}. Climate models simulate a weakening of the circulation\textsuperscript{\cite{Vallis2015, Hu2018b, Xia2020}}, driven by changes in atmospheric stability\textsuperscript{\cite{Chemke2019}}. In contrast, most reanalyses instead suggest an opposing strengthening of the NHC\textsuperscript{\cite{TANAKA2004, Mitas2005, Sohn2010, Stachnik2011, Nguyen2013a, Pikovnik2022, Zaplotnik2022b}}. This persistent disagreement has become a major source of uncertainty in assessments of large-scale atmospheric circulation change\textsuperscript{\cite{ipccar6wg1}} and has raised concerns about the credibility of both reanalysis-based trend estimates and climate model projections, undermining confidence in projected changes in tropical and subtropical precipitation. Similar disagreements have been identified for the Walker circulation, where reanalyses indicate intensification while climate models project a weakening\textsuperscript{\cite{Seager2019, Chung2019, Kosovelj2023, Watanabe2024, Toda2024}}. The recurrence of this paradox across distinct tropical overturning circulations demands a rigorous validation of circulation trends in climate models and reanalyses against direct atmospheric observations\textsuperscript{\cite{diGirolamo2025}}. 

Recent work has shown that surface pressure gradients between the tropics and subtropics provide indirect proxies for NHC strength\textsuperscript{\cite{Chemke2023}}, yielding closer agreement between climate models and observational datasets than with most reanalyses. However, surface-based diagnostics cannot capture the full three-dimensional structure of the Hadley circulation. The strength of the Hadley circulation is fundamentally defined by the meridional transport of atmospheric mass, which is quantified by the vertically integrated meridional wind throughout the troposphere\textsuperscript{\cite{Peixoto1992PhysicsClimate}}. Critically, reanalysis estimates of this quantity depend on the assimilated observations from the evolving global observing system, which may introduce spurious long-term signals. A definitive observational assessment therefore requires a constraint that is both three-dimensional and temporally homogeneous.

Here, we address this challenge using the global radiosonde record, the longest, most stable, and accurate observational dataset of atmospheric vertical profiles. To overcome the sparse and inhomogeneous distribution of radiosonde observations, we reconstruct global meridional wind fields with a masked autoencoder (MAE)\textsuperscript{\cite{He2022}} graph neural network (GNN)\textsuperscript{\cite{Brody2021}} trained on atmospheric reanalyses and applied across the radiosonde record. The MAE--GNN is trained to infer winds at unobserved locations from winds at observed locations within the same time instance by learning atmospheric teleconnections and flow-dependent correlations. When applied to sparse input data, either to radiosonde observations or their reanalysis equivalents, it uses these learned spatial relationships to reconstruct the global fields. We produce directly comparable reconstructed datasets: an observation-based reconstruction from radiosonde measurements and reanalysis-based reconstructions obtained by applying the same mapping to reanalysis values sampled at the radiosonde locations. This enables a consistent comparison of five leading reanalysis products against the observation-based reconstruction. 

Our results show that modern reanalyses systematically underestimate the climatological strength of the circulation and provide robust observational evidence that the Northern Hadley cell has weakened since 1980. By reconciling the historical radiosonde record with climate model simulations, our findings resolve a long-standing contradiction in tropical circulation trends and strengthen confidence in projections of future tropical climate shifts.

\section*{Results}

\subsection*{Systematic underestimation of NHC strength in modern reanalyses}

\begin{figure}[b!]
\centering
\includegraphics[width=\textwidth]{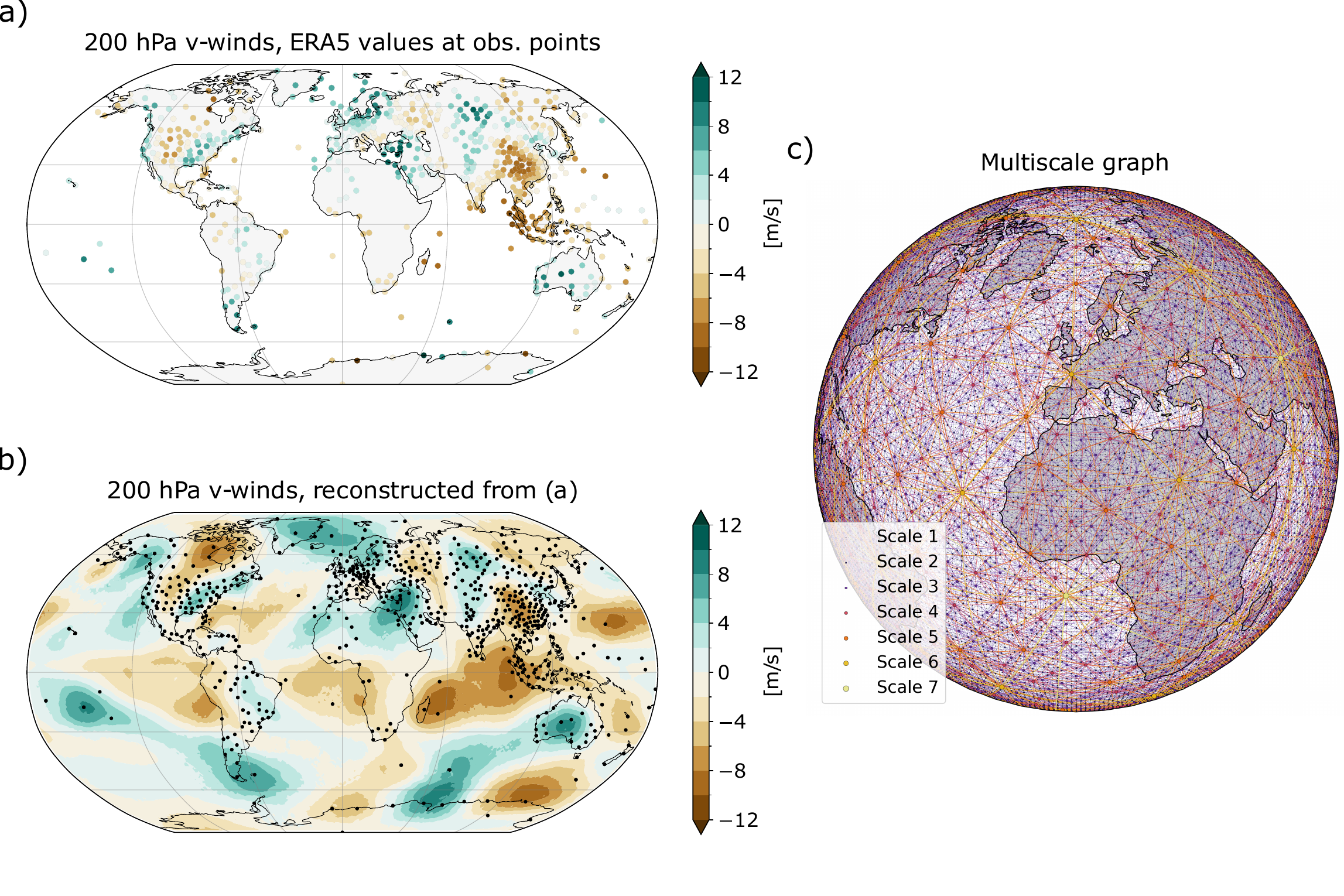}
\caption{\textbf{Reconstruction of monthly-mean meridional winds using a masked autoencoder graph neural network (MAE--GNN).} \textbf{a}, Reanalysis meridional winds sampled at radiosonde locations, serving as the sparse input to the MAE--GNN. \textbf{b}, Reconstructed meridional winds on the N80 reduced Gaussian grid, generated by the MAE--GNN from the sparse inputs in \textbf{a}. In \textbf{a} and \textbf{b}, the figure illustrates ERA5 reanalysis at 200~hPa for July 2024, with coloured/black dots indicating the locations of the radiosonde network. \textbf{c}, The multiscale graph architecture featuring graph attention mechanism\textsuperscript{\cite{Brody2021}}, representing seven distinct spatial scales for transfer of meridional wind information. Scale 1 (black nodes, $n=35,718$) represents the N80 grid with edges to 12 nearest neighbours (grey). Scales 2–7 (blue, violet, red, orange, yellow, and light yellow) denote progressively coarser grids, each containing approximately one-quarter the nodes of the preceding level, culminating in the coarsest grid of only 8 nodes (light yellow).}
\label{fig:composite}
\end{figure}

Radiosonde observations are heterogeneously distributed across the globe, with the highest densities concentrated over Northern Hemisphere landmasses (Fig.~\ref{fig:composite}). Consequently, while the Southern Hadley cell (SHC) remains sparsely observed, the Northern Hadley cell (NHC) is sufficiently sampled to permit an accurate reconstruction of its three-dimensional meridional wind field, provided that data-sparse regions such as the Atlantic and the eastern Pacific (Extended~Data~Fig.~\ref{extfig:1}) are effectively bridged. To overcome these observational gaps, we apply MAE--GNN to radiosonde observations (Methods; Fig.~\ref{fig:composite}c). 
The resulting global reconstruction maintains physical fidelity and provides spatially complete meridional wind fields that serve as our primary observational benchmark. Conceptually, this reconstruction framework is analogous to data assimilation in numerical weather prediction, enabling recovery of large-scale atmospheric features from discrete, irregularly spaced inputs\textsuperscript{\cite{Cheng2023, Thomas2024}} (Extended~Data~Fig.~\ref{extfig:2}).

To establish this reconstruction framework, the MAE--GNN is trained to map monthly-mean reanalysis values sampled at the radiosonde observation locations (Fig.~\ref{fig:composite}a) onto a regular N80 reduced Gaussian grid (Fig.~\ref{fig:composite}b). For the results presented here, training, validation and testing of MAE--GNN use ERA5 data\textsuperscript{\cite{Hersbach2020}} over the 1979--2024 period (Methods). An ensemble of 150 MAE--GNN models is trained, from which a subset is selected that most accurately reproduces the vertical structure and closure of the Northern Hadley cell (Methods). The resulting ensemble of mappings is then applied to International Global Radiosonde Archive (IGRAv2) observations\textsuperscript{\cite{Durre2018}} to reconstruct observation-determined global meridional wind fields (MAE--OBS). To ensure consistent comparison, the same MAE--GNN mappings are also applied to reanalysis values sampled at identical observation locations. Differences between these reconstructions therefore reflect genuine circulation differences rather than reconstruction artefacts.

\begin{figure}[b!]
\centering
\includegraphics[width=\textwidth]{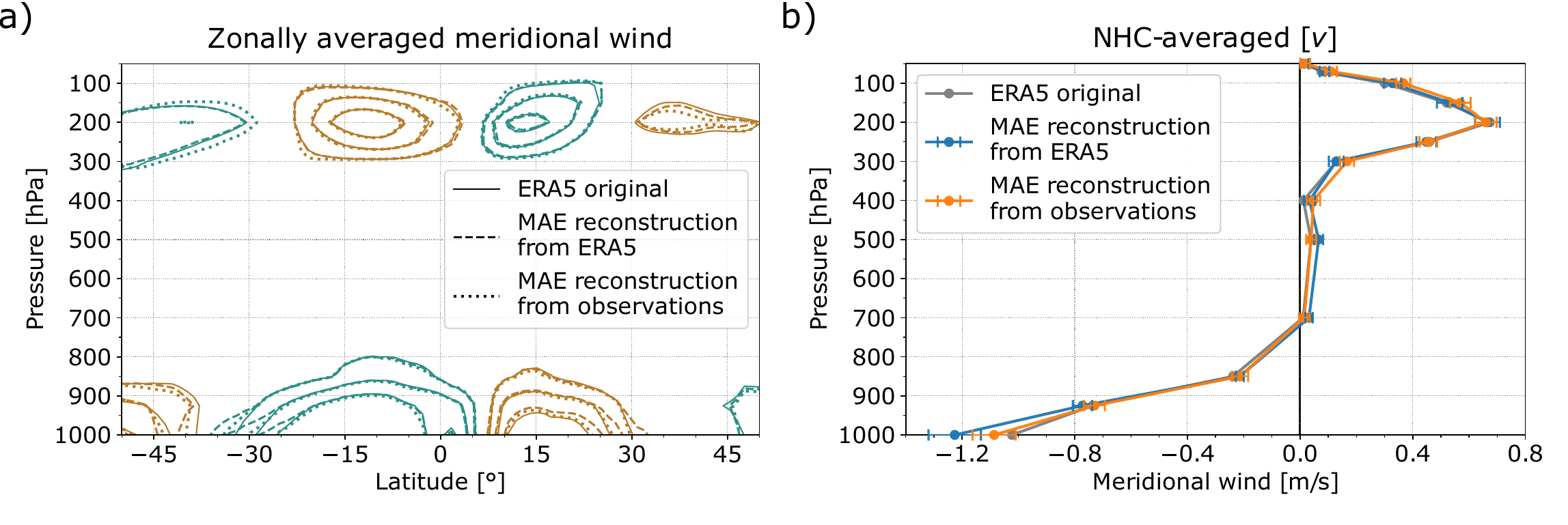}
\caption{\textbf{Evaluation of reconstructed meridional winds against ERA5 reference data.} \textbf{a}, Latitude--pressure cross-sections of the zonally-averaged meridional wind for 1980--2024. The MAE reconstruction using reanalysis values sampled at observation locations (dashed) and the reconstruction from direct radiosonde records (dotted) exhibit high spatial alignment with the original ERA5 reanalysis (solid). Green and brown contours indicate northward and southward flow, respectively (0.4 m s$^{-1}$ intervals). \textbf{b}, Vertical profile of meridional wind averaged over the region corresponding to the Northern Hadley cell (6$^\circ$N--31$^\circ$N). The reconstructed profiles (blue) closely match the vertical structure of the raw reanalysis (grey), demonstrating that the large-scale circulation is effectively captured from sparse measurements. Error bars represent the standard deviation across the ensemble of MAE reconstructions.}
\label{fig:v_comparison}
\end{figure}

This reconstruction framework enables an assessment of the time-mean structure of the Hadley circulation. The annual-mean circulation spans from 32$^\circ$S to 31$^\circ$N, with the boundary between the NHC and SHC located near 6$^\circ$N, corresponding to the mean ITCZ position\textsuperscript{\cite{Pikovnik2022}}. Each cell consists of two horizontal branches: upper-tropospheric poleward flow and lower-tropospheric equatorward flow (Fig.~\ref{fig:v_comparison}a). 

Reconstructing meridional winds using ERA5 values at observation locations successfully reproduces ERA5 circulation structure (Fig.~\ref{fig:v_comparison}a). In NHC, only a modest enhancement of the surface branch is evident in the reconstruction (Fig.~\ref{fig:v_comparison}b). 
MAE--OBS closely resembles the ERA5-based reconstruction, with small magnitude ($\le$ 0.1 m s$^{-1}$) differences. In particular, the MAE--OBS exhibits a vertically deeper and more intense upper-tropospheric poleward flow (70--400~hPa) and a weaker lower-tropospheric return flow at 1000~hPa.

\begin{figure}[b!]
\centering
\includegraphics[width=\textwidth]{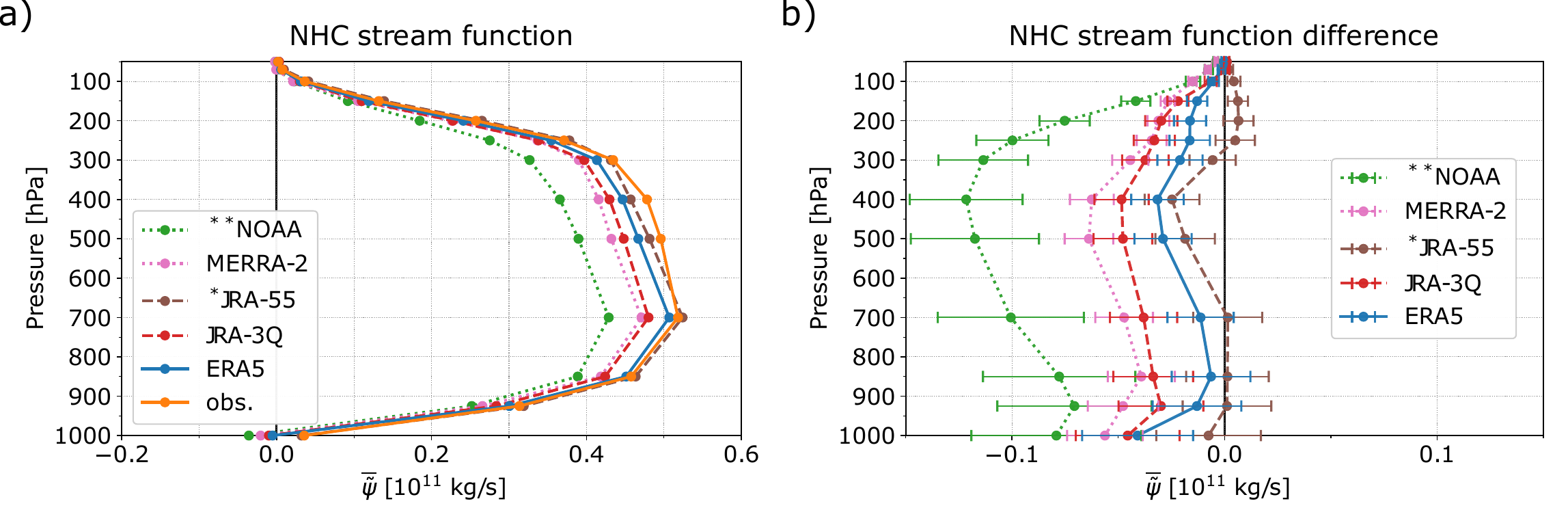}
\caption{\textbf{Discrepancies in Northern Hadley cell (NHC) strength between reanalyses-based reconstructions and radiosonde-based reconstructions (MAE--OBS).} 
\textbf{a}, Mean stream-function profiles representing climatological NHC strength for reanalyses-based reconstructions and MAE--OBS over the 1980--2024 period ($^*$JRA-55 over 1980--2023; $^{**}$NOAA 20CRv3 over 1980--2015). Profiles are derived from reconstructed meridional wind fields. 
\textbf{b}, Differences in stream-function profiles between the reanalyses and the radiosonde benchmark presented in \textbf{a}. Negative values between 400 and 700~hPa (the level of peak NHC intensity) indicate that the circulation is stronger in the MAE--OBS. Error bars denote the 95\% confidence interval derived from the paired sample t-test; intervals that do not intersect the zero line indicate a statistically significant bias ($p<0.05$).}
\label{fig:v_comparisonxx}
\end{figure}

Using MAE--OBS as a benchmark, we verify five modern reanalyses (ERA5, JRA-55\textsuperscript{\cite{KOBAYASHI2015}}, JRA-3Q\textsuperscript{\cite{Kosaka2024}}, MERRA-2\textsuperscript{\cite{Gelaro2017}} and NOAA~20CRv3\textsuperscript{\cite{Silvinski2021}}) by applying the same reconstruction mapping separately to each reanalysis, using values sampled at the observation locations. Among the products examined, the upper-tropospheric poleward flow is strongest and verifies best with observations in JRA-55 and ERA5, whereas the surface-observations-based NOAA 20CRv3 reanalysis substantially underestimates it, indicating that surface observations alone cannot properly constrain the Hadley circulation (Extended~Data~Fig.~\ref{extfig:3}). These differences are amplified when the NHC strength is represented with a stream function (Fig.~\ref{fig:v_comparisonxx}a; Methods, Eq.~\ref{eq:hc_here}), revealing a systematic underestimation of peak NHC strength (400--700~hPa) across all reanalysis products (Fig.~\ref{fig:v_comparisonxx}b). JRA-55 and ERA5 exhibit the smallest biases, NOAA 20CRv3 shows the largest bias (approximately 20\% underestimation), and JRA-3Q and MERRA-2 underestimate NHC strength by about 10--15\%. These results are insensitive to the choice of training reanalysis: the reconstructed circulation remains qualitatively and quantitatively consistent when the MAE--GNN is trained on JRA-3Q, \mbox{JRA-55}, or MERRA-2 (Extended~Data~Fig.~\ref{extfig:4}). While the JRA-55 reconstruction exhibits a modest dependence on the training dataset by exceeding the radiosonde-derived NHC strength in the lower troposphere when trained on \mbox{JRA-3Q}, the systematic underestimation across the ensemble remains the dominant and defining feature.

Importantly, the relative differences in NHC strength among the reanalysis-based reconstructions are consistent with those inferred directly from the raw reanalyses (Extended~Data~Fig.~\ref{extfig:5}a).
This consistency reinforces the validity of our results, confirming that modern reanalyses, regardless of their underlying model or data assimilation scheme, systematically underestimate the magnitude of the observed Hadley circulation.

\subsection*{Modulation of circulation strength by data assimilation increments}

\begin{figure}[t!]
\centering
\includegraphics[width=\textwidth]{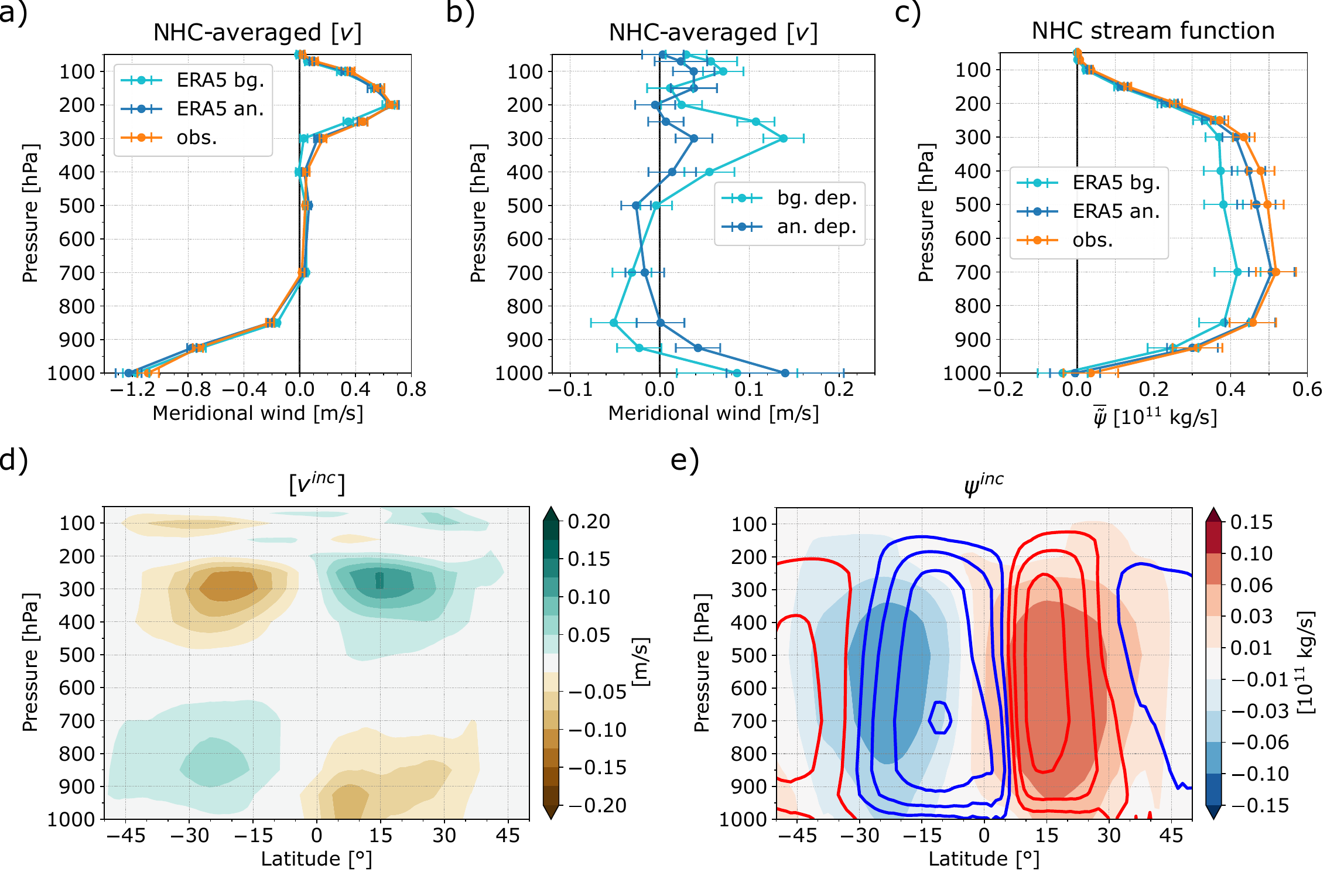}
\caption{\textbf{Data assimilation systematically strengthens the Northern Hadley cell.} Profiles in \textbf{a--c} are averaged over the extent of the NHC; \textbf{d,e} show zonal-mean cross-sections. All panels represent the 1980--2024 period and are derived from reconstructed wind fields.
\textbf{a}, Mean meridional wind profiles in ERA5 background (cyan), ERA5 analysis (blue), and radiosonde observations (orange).
\textbf{b}, Mean meridional wind background departures (observations minus background, cyan) and analysis departures (observations minus analysis, blue).
\textbf{c}, Mean stream function. The ERA5 background is consistently corrected (ERA5 analysis) towards a stronger radiosonde-observed circulation, thereby correcting the mean state. The discrepancy between the background and observations highlights the model's intrinsic underestimation of NHC strength.
\textbf{d}, Zonal-mean meridional wind analysis increments. The quadrupole pattern illustrates how the corrections from the global observing system intensify both the Northern and Southern Hadley cells.
\textbf{e}, Climatological stream function (contours) and analysis increments (colours). Contours indicate the climatological Hadley circulation strength, with red contours for positive stream-function values (0.1, 0.3, 0.6, 1 $\times$ 10$^{11}$ kg s$^{-1}$) and blue for negative counterparts. Error bars in \textbf{a--c} denote the standard deviation across the ensemble.}
\label{fig:v_inc_comparison}
\end{figure}

Reanalyses estimate past atmospheric states by optimally correcting short-range model forecasts (the background state) with observations through data assimilation. 
To assess whether the systematic underestimation of NHC strength in reanalyses originates from the data assimilation process, we compare reconstructed NHC meridional winds from the ERA5 background, ERA5 analysis, and radiosonde observations. The background underestimates both the upper-tropospheric poleward flow and the lower-tropospheric equatorward flow at the 850 and 925~hPa pressure levels (Fig.~\ref{fig:v_inc_comparison}a). This bias is also evident in the background departures (observations minus background; Methods) of NHC meridional winds (Fig.~\ref{fig:v_inc_comparison}b) and in the corresponding stream-function profiles (Fig.~\ref{fig:v_inc_comparison}c). 

These systematic biases are partly mitigated by the data assimilation system, which applies a mean-state correction to the model's circulation. This correction is visible as a quadrupole pattern of meridional wind analysis increments (analysis minus background; Fig.~\ref{fig:v_inc_comparison}d) and associated stream-function increments in the latitude--pressure plane (Fig.~\ref{fig:v_inc_comparison}e), indicating a systematic intensification of the overturning circulation in both hemispheres. The correction pattern closely resembles the increments diagnosed from the raw ERA5 data (Extended~Data~Fig.~\ref{extfig:6}). The increments also act to widen the NHC and shift the zonal-mean ITCZ southwards (Fig.~\ref{fig:v_inc_comparison}e, Extended~Data~Fig.~\ref{extfig:6}d). Although analysis departures are reduced relative to background departures (Fig.~\ref{fig:v_inc_comparison}b), the reanalysis NHC remains weaker than in the radiosonde-based reconstructions (Fig.~\ref{fig:v_inc_comparison}a,c). The exception is the 1000~hPa level, where analysis departures exceed background departures, indicating that DA drives the near-surface meridional winds away from observations (Fig.~\ref{fig:v_inc_comparison}b).

Taken together, our results indicate a systematic cycle: DA strengthens the Hadley circulation at each assimilation step, after which short-range forecast integration relaxes the circulation toward a weaker model climatology, before being corrected again by DA. The persistence of this correction–relaxation cycle throughout the reanalysis period indicates an intrinsic forecast-model bias rather than a deficiency in the global observing system or the DA procedure. Because data assimilation can, by construction, only partially close the gap to observations, this bias is carried into the resulting ERA5 reanalysis. This is consistent with the systematic underestimation of NHC strength in reanalyses identified above.

\subsection*{Emergence of a weakening trend in the stable observational record}

\begin{figure}[b!]
\centering
\includegraphics[width=\textwidth]{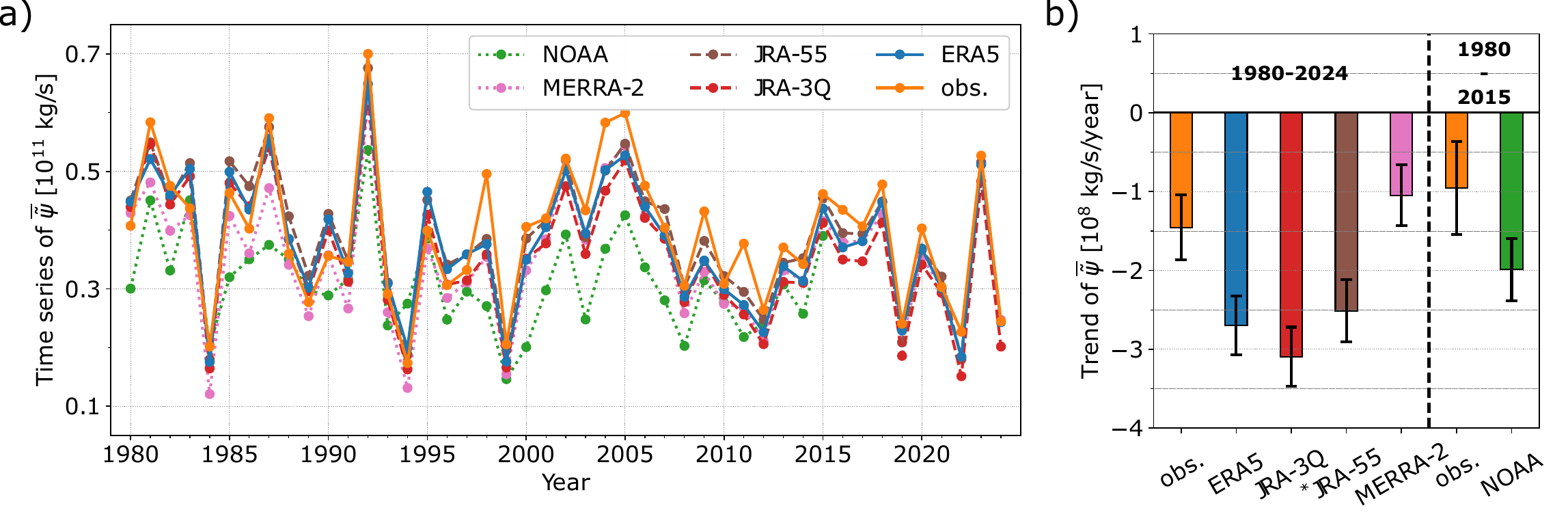}
\caption{\textbf{Observed weakening of the Northern Hadley cell and associated trends.} 
\textbf{a}, Time series of the annual-mean stream-function strength $\overline{\tilde{\psi}}$. Reanalysis-based reconstructions and MAE--OBS show high correlation in interannual variability.
\textbf{b}, Multidecadal trends in NHC strength for the 1980--2024 period ($^*$JRA-55 over 1980--2023) and the 1980--2015 period. The observational reconstruction (orange) and all reanalysis-based reconstructions show a consistent weakening. Error bars denote the 95\% confidence intervals.}
\label{fig:trends}
\end{figure}

To quantify long-term changes in NHC strength, we use a single-value stream-function-based metric (Methods). Time series derived from this metric show strong coherence in interannual variability between reconstructions based on radiosonde observations and those derived from reanalyses over the 1980--2024 period (Fig.~\ref{fig:trends}a). 

Both MAE--OBS and reanalysis-based reconstructions reveal a statistically significant weakening of the NHC since 1980 (Fig.~\ref{fig:trends}b). The weakening trends inferred from ERA5, JRA-3Q, and JRA-55 exceed the magnitude of the radiosonde-based estimate, whereas MERRA-2 aligns more closely with observations. Despite the differences in magnitude, this result provides critical observational confirmation of weakening trends in NHC strength, consistent with long-standing projections from climate model historical simulations\textsuperscript{\cite{Chemke2019}}. The robustness of this trend is confirmed across a wide range of sensitivity tests. The trend remains statistically significant regardless of the reanalysis used for MAE--GNN training (ERA5, JRA-3Q, JRA-55, MERRA-2) or the specific metric employed to define circulation strength (Extended~Data~Fig.~\ref{extfig:7}). Furthermore, the weakening persists when the trend is evaluated over different time intervals (Extended~Data~Fig.~\ref{extfig:8}). The weakening trend remains statistically significant, with little change in magnitude when 10\%, 20\% or 40\% of radiosonde stations are removed from the observing network in observation-denial experiments that preserve network's large-scale spatial structure (Extended~Data~Fig.~\ref{extfig:9}; Methods). This invariance strongly suggests that the signal reflects a persistent physical change rather than an artefact of the reconstruction methodology.

In contrast to MAE--GNN reconstructions, among the raw reanalysis products, a statistically significant weakening emerges only in ERA5 (Extended~Data~Fig.~\ref{extfig:5}c,d), first becoming detectable in the 1980--2021 period and persisting thereafter (Extended~Data~Fig.~\ref{extfig:8}). This discrepancy suggests that the weakening signal is consistently captured by the reconstructions but remains muted or inconsistent in the raw reanalyses. A likely explanation is the strong temporal inhomogeneity of the global observing system: modern reanalyses assimilate a volume of observations that has increased by orders of magnitude over recent decades\textsuperscript{\cite{Hersbach2020}}, introducing artificial shifts in circulation strength. Instead, our reconstruction approach applies a time-invariant MAE--GNN mapping from radiosonde measurements (or their reanalysis equivalents) to the full global grid. Because the radiosonde network is comparatively more stable over time relative to the rapidly evolving global observing system, and because this mapping is fixed across the entire study period, it provides a temporally homogeneous framework for estimating long-term circulation changes, enabling robust detection of multidecadal trends. While individual radiosonde station data exhibit known discontinuities due to instrument changes and station relocations, the NHC metric aggregates over a sufficiently large Northern Hemisphere network that station-level inhomogeneities are unlikely to systematically bias the zonal-mean circulation estimate\textsuperscript{\cite{Durre2018}}. Moreover, homogenization studies show that systematic wind adjustments at individual radiosonde stations are relatively small\textsuperscript{\cite{Madonna2022}}. The robustness of the weakening trend across all training reanalyses, circulation metrics, and analysis periods provides further empirical confirmation that the signal is not an artefact of either the observational network or the reconstruction methodology.

Previous studies have proposed that data assimilation artefacts may induce spurious strengthening trends in reanalyses\textsuperscript{\cite{Chemke2019, Chemke2023}}. We find that the magnitude of annual-mean stream-function increments in ERA5 (the systematic corrections identified in the previous sections) has decreased markedly over recent decades (Extended~Data~Fig.~\ref{extfig:10}). As the observing system has expanded, background forecasts have improved, requiring progressively smaller corrections. 
Because these increments act to intensify the Hadley circulation, their diminishing magnitude over time reduces the DA-driven strengthening, thereby amplifying a weakening tendency in the reanalysis-based NHC trend (Extended~Data~Fig.~\ref{extfig:10}). Our results thus demonstrate that, while DA influences NHC trends in ERA5, the resulting increments have paradoxically contributed to a weakening rather than the spurious strengthening trend previously attributed to DA artefacts\textsuperscript{\cite{Chemke2019}}.

\section*{Discussion}

Our study provides an observation-based assessment of recent multidecadal changes in Hadley circulation strength using radiosonde measurements. By reconstructing the three-dimensional meridional wind field from sparse observations with a novel masked autoencoder (MAE) graph neural network (GNN), we demonstrate that a temporally stable radiosonde network can constrain the large-scale circulation despite geographic sampling inhomogeneity. 

Applying this reconstruction framework shows that the strength of the Northern Hadley cell (NHC) is systematically underestimated in modern atmospheric reanalyses relative to radiosonde observations. The bias is most pronounced in the surface-observations-based NOAA 20CRv3 reanalysis, whereas ERA5 and especially JRA-55 exhibit smaller discrepancies. This highlights the limited ability of surface-only reanalyses to constrain the full vertical structure of the Hadley circulation. A potential concern is that MAE--GNN training on reanalysis data could introduce reanalysis biases into reconstructions based on radiosonde observations (MAE--OBS). However, this is inconsistent with our results: if our reconstruction model had inherited reanalysis biases, MAE--OBS would be expected to show similar NHC strength as reanalyses. Instead, MAE--OBS systematically exceeds reanalysis-based reconstructions, and this finding is insensitive to the choice of training reanalysis. Together, these results indicate that MAE--OBS reflects the radiosonde observations rather than the climatology of the training dataset.

In ERA5, data assimilation (DA) increments systematically enhance both the Northern and Southern Hadley cells, indicating that the forecast model underestimates their amplitudes. These findings are consistent with evidence that ERA5 DA enhances surface meridional winds over the Pacific\textsuperscript{\cite{Mayer2025a}} and that optimal initial-condition perturbations for improving weather predictability involve large-scale corrections that strengthen the Hadley circulation\textsuperscript{\cite{Vonich2025}}.

The persistent underestimation of NHC strength in the ERA5 forecast model likely reflects a combination of structural physical model deficiencies, including a boundary-layer dry bias\textsuperscript{\cite{Hersbach2020}} and limited horizontal resolution (31 km). Although future higher-resolution reanalyses may reduce these mean-state biases, our results demonstrate that DA currently performs a critical, albeit partial, correction of the large-scale circulation. Because DA, by definition, can only adjust the analysed state toward observations while remaining constrained by the short-range forecast model, it cannot fully eliminate systematic model biases. The persistent requirement for such corrections therefore provides an independent line of evidence that NHC strength remains underestimated in the resulting ERA5 reanalysis.
Beyond direct wind adjustments, DA can also modulate the Hadley circulation indirectly, for example, by correcting humidity profiles that influence latent heat release and convective intensity in the cycled short-range forecast\textsuperscript{\cite{McNally1996}}. Understanding the net impact of these corrections on large‑scale circulation remains an important subject for future investigation.

Importantly, our analysis provides the first vertically resolved observational evidence that the Northern Hadley cell has weakened since 1980. Verifying changes across the full vertical structure of the circulation is essential, as illustrated by the inability of surface-only NOAA 20CRv3 reanalysis to constrain the mid- to upper-tropospheric flow. The detected weakening is supported by two key lines of evidence: first, consistent weakening trends across MAE--OBS and all reanalysis-based reconstructions, and second, the emergence of a statistically significant weakening trend in raw ERA5 reanalysis, first becoming detectable in the 1980--2021 period. Earlier studies using shorter reanalysis records therefore understandably reported strengthening\textsuperscript{\cite{TANAKA2004, Mitas2005, Sohn2010, Stachnik2011, Nguyen2013a, Chemke2019, Pikovnik2022, Zaplotnik2022b}}.

The discrepancy in trends between raw reanalyses and reconstructions likely arises from the evolving global observing system. As reanalyses assimilate an increasing volume of observations, spurious discontinuities can mask or distort underlying climate signals. In contrast, our reconstruction applies a fixed, time-invariant nonlinear mapping from a relatively stable set of radiosonde locations to the full global grid, providing a temporally homogeneous framework for diagnosing long‑term circulation change. More broadly, the MAE--GNN provides a tool for evaluating climate trends in the presence of a changing global observing system.

Overall, we provide compelling evidence for the recent weakening of the Northern Hadley cell, consistent with historical climate model simulations. This agreement strengthens confidence in future projections of Hadley circulation change and its implications for global precipitation patterns.

\section*{Methods}

\subsection*{Reanalysis and observation data}

This study uses several atmospheric reanalyses to investigate the Northern Hadley cell strength: ERA5\textsuperscript{\cite{Hersbach2020}} from ECMWF, JRA-55\textsuperscript{\cite{KOBAYASHI2015}} and JRA-3Q\textsuperscript{\cite{Kosaka2024}} from JMA, \mbox{MERRA-2}\textsuperscript{\cite{Gelaro2017}} from NASA, and NOAA 20CRv3\textsuperscript{\cite{Silvinski2021}}. Among these, ERA5 provides the highest spatial resolution (31 km, 137 model levels), most detailed physics, and largest volume of assimilated data, leading to improved agreement with observations for tropospheric temperature, wind, humidity, and precipitation\textsuperscript{\cite{Simmons2022}}. ERA5 also notably reduces biases in surface meridional wind and horizontal wind divergence over oceans\textsuperscript{\cite{Rivas2019}}, and its atmospheric energy, moisture, and mass budgets are demonstrably superior to previous-generation reanalyses like ERA-Interim\textsuperscript{\cite{Mayer2021}}.

JRA-55 (55 km resolution, 60 levels) assimilates extensive conventional and satellite observations but exhibits known upper-tropospheric warm bias and excessive tropical precipitation\textsuperscript{\cite{KOBAYASHI2015, Harada2016}}. JRA-3Q (40 km horizontal resolution, 100 vertical layers), the successor to JRA‑55, incorporates advances from the operational NWP system since JRA-55 and improves the global energy budget\textsuperscript{\cite{Kosaka2024}}.

MERRA-2 (50 km resolution, 72 levels) shows a significant positive temperature bias in the mid-to-upper troposphere and excessive precipitation in regions with high tropical topography\textsuperscript{\cite{Gelaro2017,fujiwara2022}}.

To isolate the impact of upper-air observations on Hadley circulation strength, we also analyse NOAA 20CRv3 (75 km resolution, 64 levels), which assimilates only surface pressure observations and relies on prescribed sea‑surface temperature and sea ice concentration boundary conditions, deliberately omitting satellite and upper-air measurements.  Despite known biases in temperature and wind above 300~hPa, its long time series and multidecadal means of mass, circulation, and precipitation generally align with other modern reanalyses and observation-based products\textsuperscript{\cite{Silvinski2021}}, providing a useful baseline for assessing the role of upper‑air data.

We obtain the meridional wind ($v$) data from reanalyses at standard pressure levels: 37 pressure levels in ERA5 and JRA-55, 45 levels in JRA-3Q, 42 levels in MERRA-2, and 28 levels in NOAA 20CRv3; and on a latitude-longitude grid with 1° resolution in ERA5 and NOAA 20CRv3, 1.25° in JRA-3Q and JRA-55, and 0.5° x 0.625° in MERRA-2. For consistency and further use in a masked autoencoder, all reanalysis data were post-processed to an N80 reduced Gaussian grid\textsuperscript{\cite{ecmwf_N80}}. The analysis times are set at 00 and 12~UTC, and their monthly means are compared to the corresponding monthly-mean radiosonde observations.

Our observational dataset comprises quality-controlled radiosonde data from the Integrated Global Radiosonde Archive Version 2.2 (IGRAv2)\textsuperscript{\cite{Durre2018}}, distributed by NOAA. IGRAv2 data are not adjusted for inhomogeneities due to station relocation, changes in instrumentation, or changes in observing practice. We use monthly-mean meridional wind values, which are available at 21 standard pressure levels (1000 to 1~hPa), computed at 00~UTC and 12~UTC, using data collected within 2 hours of each nominal time. The IGRAv2 dataset includes specific years/months with a sufficient number of standard-pressure-level observations at fixed land-based stations (i.e., it does not account for the drift of radiosonde balloons). The vertical distribution of radiosonde observations is relatively uniform throughout the troposphere, except at 925~hPa and 1000~hPa, which lie beneath the surface in many regions. Observation locations are unevenly distributed between the hemispheres. The NHC, the focus of this study, is much better observed than the Southern Hadley cell. 

To perform a consistent comparison between reanalyses and observations, we use an intersection of 13 pressure levels for both reanalyses and observational datasets: 1000, 925, 850, 700, 500, 400, 300, 250, 200, 150, 100, 70, and 50~hPa.

Our analysis spans 45 years of monthly-mean radiosonde data (1980--2024) and up to 45 years of reanalysis data (1980--2024 for ERA5, JRA-3Q, MERRA-2; 1980--2023 for JRA-55; 1980--2015 for NOAA 20CRv3). Throughout this period, the distribution of radiosonde data is rather homogeneous, both globally as well as in the Hadley cell region, with a slight drop in 2020 and 2021, likely due to COVID-19-related measures. The seasonal distribution of observations is uniform, with comparable numbers of radiosonde observations in each calendar month. The properties of the radiosonde data applied in this study are further depicted in Supplementary~Information~Figs.~1--4, while the features of the climatological Hadley circulation are illustrated in Supplementary~Information~Figs.~5,~6.

\subsection*{ERA5 background and analysis increments}

Data assimilation (DA) is a cyclical process that objectively combines a short-range forecast (background, $\mathbf{x}_b$) from the previous DA cycle with available Earth-system observations ($\mathbf{y}$) accounting for their respective error characteristics, to produce an optimal estimate of the atmospheric state at a given time, known as the analysis ($\mathbf{x}_a$). The analysis is then used to initialise the next short-range forecast, and the cycle repeats. A sequence of such analyses over a historical period produces a reanalysis. 

The correction applied to the background state at each cycle, the analysis increment, is given by
\begin{equation}
    \delta\mathbf{x}_a = \mathbf{x}_a - \mathbf{x}_b \, .
\end{equation}
Here, the state vector $\mathbf{x}$ represents meridional wind fields at a specific time, stacked into a single vector.

To investigate how DA increments influence the representation of the Hadley circulation in reanalyses, we use background data from ERA5, which provides publicly accessible background fields. In ERA5, the analysis increments are computed cyclically using 4D-Var\textsuperscript{\cite{Courtier1994}} DA that uses 12-hour long assimilation windows\textsuperscript{\cite{Hersbach2020}}. Therefore, the monthly-mean analysis increments were taken as the difference between the monthly-mean analysis values ($\mathbf{x}_a$) valid at 00/12~UTC and the monthly-mean of the previous short-range (6-hour) forecast ($\mathbf{x}_b$) valid at the same time and initialised at 18/06~UTC.

\subsection*{Comparison of Meridional Winds in Reanalysis and Radiosonde Observations}

The reanalysis meridional winds are provided on a regular grid, whereas radiosonde observations are sparse and irregularly distributed. These datasets can be directly compared either in the observation space by mapping the reanalyses to observation locations, or in the grid-point space by reconstructing full global fields from sparse observations.

Reanalyses and irregular observations are directly compared by computing analysis departures ($\mathbf{d}_a$) in observation space -- a standard approach in data assimilation for numerical weather prediction\textsuperscript{\cite{Lahoz2013}}. Analysis departures are defined by contrasting radiosonde observations ($\mathbf{y}$) with analysis values mapped to the observation space using the observation operator $\mathcal{H}$:
\begin{equation}
    \mathbf{d}_a=\mathbf{y}-\mathcal{H}(\mathbf{x}_a) \, .
\end{equation}
As the radiosonde-measured meridional wind directly matches the reanalysis state variable, the observation operator $\mathcal{H}$ simply represents bilinear interpolations of analysis values to the observation locations. 
Similarly, the background departures ($\mathbf{d}_b$) quantify the difference between observations and the background:
\begin{equation}
    \mathbf{d}_b=\mathbf{y}-\mathcal{H}(\mathbf{x}_b) \, .
\end{equation}
In this study, $\mathbf{x}_a$, $\mathbf{x}_b$,  $\mathbf{y}$ and the derived vectors ($\delta\mathbf{x}_a$, $\mathbf{d}_a$, $\mathbf{d}_b$) represent monthly-mean values of meridional wind.

A limitation of observation-space comparison is that the spatial inhomogeneity of the radiosonde observations prevents an accurate calculation of the zonal-mean meridional winds required to quantify the Hadley circulation strength. To represent the Hadley circulation using sparse and inhomogeneous radiosonde data (or their reanalyses equivalents), we successively evaluated three distinct averaging approaches:
\begin{enumerate}
    \item \textbf{Binned averaging}: Partitioning data into six longitudinal bins of 60$^\circ$ width, computing averages within each bin, and then averaging across bins.
    \item \textbf{Spherical nearest-neighbour averaging}: Performing spherical nearest-neighbour interpolation from observation space to the full grid, followed by averaging in model space.
    \item \textbf{Masked autoencoder (MAE) approach}: Applying a deep learning model based on a graph neural network (GNN) as a data-driven interpolator from observation space to the full global grid, followed by averaging in model space.
\end{enumerate}
When characterising the NHC, binned averaging offers only marginal improvements over naive averaging in the observation space. While the nearest-neighbour method provides a significantly more accurate representation, it fails to satisfy the mass continuity, as shown by the unclosed stream-function isolines (Extended~Data~Fig.~\ref{extfig:5}b). The MAE--GNN approach proves the most robust. By mapping sparse observations (or their reanalyses equivalents) onto a regular global grid, the GNN functions as a physically-informed interpolator that produces a complete, accurate and consistent reconstruction. A comparison of different methods for reconstructing the NHC is shown in Supplementary~Information~Fig.~7.

\subsection*{Reconstructing Global Meridional Wind Fields from Sparse Radiosonde Data - Masked Autoencoder Graph Neural Network}

MAE--GNN is trained on reanalyses to perform a data-driven interpolation from reanalysis values sampled at sparse radiosonde locations to the full global grid (Fig.~\ref{fig:composite}a,b). By learning complex nonlinear correlations between sparse points and the global atmospheric state, the neural network can reconstruct the missing (masked) part of the field from radiosonde observations. Two distinct examples of the impact of a single radiosonde measurement are shown in Supplementary~Information~Fig.~8.

The model is structured to reconstruct a stack of 2D horizontal fields, pressure level by pressure level, rather than a single 3D field, at each time instance. Such an approach is taken to increase the training set size (see section Training data). Because the levels are processed independently, the vertical coherence of the reconstructed Hadley circulation is not guaranteed by the network architecture. To ensure a physically consistent 3D structure of the Hadley circulation, we employ an ensemble-based filtering approach. We trained a large ensemble of 150 neural networks with the same architecture but different initial weights for training. We retain only networks that satisfy the fundamental structural requirements of the NHC (e.g., mass-balance closure and vertical alignment of circulation cells, see section Physical Consistency Constraint), and average the reconstructions. This post-processing step ensures that the final product maintains physically consistent three-dimensional Hadley circulation (Fig.~\ref{fig:v_comparison}). 

\subsubsection*{Training data}

The MAE--GNN was trained using 46 years (1979--2024) of monthly-mean ERA5 data at 00/12~UTC, retrieved from Copernicus Climate Change Service (C3S) Climate Data Store (CDS) (2025). The training data consists of 13 pressure levels on a reduced Gaussian grid with horizontal resolution N80, corresponding to 160 latitude circles and 320 points around the equatorial latitude circle (1.125° resolution around the equator), a total of 35718 grid points (Fig.~\ref{fig:composite}c). 

The fields were standardised using the 3D grid-point climatological mean and standard deviation:
\begin{equation}
    v^s(t,p,\phi,\lambda) = \frac{v(t,p,\phi,\lambda)-\overline{v}(p,\phi,\lambda)}{\sigma_v(p,\phi,\lambda)}\, ,
\end{equation}
where $v^s$ denotes standardised meridional wind field, $v$ is raw meridional wind, and $\overline{v}$ its 3D mean. The standard deviation $\sigma$ is computed over the full three‑dimensional field, and the normalisation uses $\sigma_v = \mathrm{max}(\sigma,\epsilon)$  with $\epsilon=0.1$ imposed to prevent training instability in regions of low variance. $p$, $\phi$, $\lambda$ denote pressure, latitude and longitude, respectively.

The use of monthly-mean data (instead of instantaneous atmospheric states) provides a smoother target and allows the MAE--GNN to better generalise the mapping from sparse location to global model state. However, this vastly decreases the size of the training dataset. To increase the size of the training data, we flattened the data across pressure levels. This yields a dataset with 14352 instances, i.e., 46 years $\times$ 12 months/year $\times$ 2 instances per month (00/12~UTC) $\times$ 13 pressure levels. This dataset was partitioned into training, validation, and testing sets with relative proportions of 0.83, 0.1, and 0.07, respectively. Thus, approximately 37 years were used for training, 6 for validation, and 3 for testing the model. While flattening increases the training size, it leaves the MAE--GNN vertically unconstrained. Therefore, as described in the continuation (section Physical Consistency Constraint), we trained an ensemble of networks to ensure that reconstructions have appropriate vertical structure.

For sensitivity testing, the MAE--GNN was also independently trained using \mbox{JRA-3Q} data (1979--2024), JRA-55 data (1979--2023), and MERRA-2 data (1980--2024). We opted not to train the MAE--GNN on NOAA 20CRv3 data, as this reanalysis fails to properly constrain upper-tropospheric flow and significantly underestimates the NHC strength.

\subsubsection*{Multiscale graph representation}

To represent the spatial domain of the MAE--GNN, we constructed a hierarchical sequence of $k$-nearest-neighbour (kNN) graphs on a spherical latitude--longitude grid, forming a multiscale cumulative graph (Fig.~\ref{fig:composite}c). The primary graph uses an N80 reduced Gaussian grid with 35718 nodes. From this finest-scale graph, we generated six progressively coarser kNN subgraphs by subsampling nodes by a factor of $4$ at each level, yielding graphs with $8930$, $2232$, $558$, $140$, $35$, and $8$ nodes at scales 2 through 7, respectively. 

Both the base graph and the subgraphs use $k=12$ nearest-neighbour connections. This multiscale graph architecture is designed to capture atmospheric features across a broad range of spatial scales, from localised correlations to global-scale information transfer. The connectivity of an example scale-6 node is illustrated in Supplementary~Information~Fig.~9.

\subsubsection*{MAE--GNN architecture}

We formulate the MAE--GNN as follows. The input tensor is of shape 35718$\times$3, corresponding to 35718 grid nodes with three features per node: meridional wind, latitude, and longitude (the latter two are static). The input tensor is passed through the encoder to a latent representation and then through the decoder, which produces an output tensor of shape 35718$\times$1 containing meridional wind only. 

To mask meridional wind values at unobserved nodes in the input tensor, we apply a time- and pressure-dependent binary mask to the meridional-wind feature. Masked entries are replaced by a single learnable mask token, which is updated during training and represents an unknown (blank) value.

To determine which nodes are observed, each observation is assigned to its nearest grid node. Avoiding interpolation operators was a necessary computational simplification, which introduces a small representativeness error, but does not meaningfully alter our results.

The encoder is structured as a five-stage progressive architecture with the following blocks: the initial projection block, three graph attention blocks, and the final latent projection block:
\begin{itemize}
    \item \textbf{Initial projection}: Node features are mapped into a 16-dimensional space with a linear transformation, followed by layer normalisation, feature dropout (rate of 0.1), and a Gaussian Error Linear Unit (GELU) activation. A residual connection bridges the input and the transformed features to improve gradient backpropagation.
    \item \textbf{Graph attention blocks}: Each block consists of a GATv2 convolution layer\textsuperscript{\cite{Brody2021}}, which implements an attention mechanism where every node $i$ computes a score $e_{ij}$ to weight the influence of adjacent nodes $j$. The graph attention layer maps 16 features to 4 features per head across 4 parallel attention heads, which are then concatenated to maintain a 16-dimensional feature vector. The attention mechanism includes self-attention, uses a learnable bias term and internal skip connections. After the attention layer, the output undergoes normalisation, a feature dropout, and GELU activation. To ensure stability and mitigate the vanishing gradient problem, an outer residual connection sums the block's input with the activated output.
    \item \textbf{Final latent projection}: It maps the features to the latent space. To mitigate information loss, we employ a multi-layer perceptron (MLP) with an expanded intermediate dimension (32 units), GELU activation, and final compression to a $16$-dimensional latent vector. 
\end{itemize}
The decoder mirrors the encoder, mapping the latent representation to the output space.

\subsubsection*{Training procedure}

The MAE is trained to map reanalysis values sampled at radiosonde locations (typically $\sim600$ data points per instance; all other nodes are masked) back to the full global grid (35718 nodes). The mapping from the sparse input $\mathcal{H}(\mathbf{x}_a)$ to the reconstructed full field  $\widetilde{\mathbf{x}}_a$ is 
\begin{equation}\label{eq:reanalysis_reconstruction}
    \widetilde{\mathbf{x}}_a = \mathcal{G}_{NN} \left(\mathcal{H}(\mathbf{x}_a)\right) \, ,
\end{equation}
where $\mathcal{G}_{NN}$ denotes the neural-network operator. Training minimises the mean-squared error (MSE) between the original reanalysis field $\mathbf{x}_a$ and the reconstruction $\widetilde{\mathbf{x}}_a$, evaluated only at masked (unobserved) nodes:
\begin{equation}
    \mathcal{L}_{\mathrm{MSE}} = \frac{1}{N}\sum_{i=1}^{N} \left\| \mathbf{m}_i \odot \left[ \left(\widetilde{\mathbf{x}}_a\right)_i - \left(\mathbf{x}_a\right)_i \right] \right\|^2 ,
\end{equation}
where $N$ is the number of training instances, $\mathbf{m}_i$ is a binary mask vector (1 at masked nodes, 0 otherwise), and $\odot$ denotes elementwise multiplication. The mask is time- and level-dependent because the horizontal distribution of radiosonde observations varies with time and pressure level.

For loss function minimisation, we used the AdamW optimizer with a weight decay of $1 \times 10^{-4}$ to prevent overfitting. The learning rate was controlled by a Cosine Annealing scheduler, enabling rapid initial convergence followed by fine-tuning during later training stages. To manage memory constraints, we applied gradient checkpointing in the middle GAT layers, recomputing activations during the backward pass rather than storing them, which enables deeper graph models within GPU limits.

\subsubsection*{Reconstructions}

Once trained, the MAE--GNN is applied to the IGRAv2 radiosonde record to generate full-grid reconstructions from real observations. The reconstruction $\widetilde{\mathbf{x}}_y$ is derived from the observation vector $\mathbf{y}$ as
\begin{equation}
   \widetilde{\mathbf{x}}_y = \mathcal{G}_{NN} \left(\mathbf{y}\right) .
\end{equation}
Extended~Data~Fig.~\ref{extfig:2} shows examples of reconstructing a full-grid field from radiosonde observations in the upper troposphere (Extended~Data~Fig.~\ref{extfig:2}c) and lower troposphere (Extended~Data~Fig.~\ref{extfig:2}f). 

For consistent comparison between reanalyses and observations, we evaluate the reanalysis reconstructions $\widetilde{\mathbf{x}}_a$ (Eq.~\ref{eq:reanalysis_reconstruction}) directly against the radiosonde reconstructions $\widetilde{\mathbf{x}}_y$. By producing both datasets with the same $\mathcal{G}_{NN}$ operator, we effectively match the spatial representations of the two products. This approach ensures that any reconstruction biases arising from the sparse sampling of the radiosonde network are applied consistently to both datasets, allowing departures $\widetilde{\mathbf{d}}_a=\widetilde{\mathbf{x}}_y - \widetilde{\mathbf{x}}_a$ and $\widetilde{\mathbf{d}}_b= \widetilde{\mathbf{x}}_y - \widetilde{\mathbf{x}}_b$ to be attributed to genuine differences in the monthly-mean atmospheric state rather than sampling artefacts.
The analysis increments are therefore computed consistently as the difference between the reconstructed analysis and background fields:
\begin{equation}
    \widetilde{\delta\mathbf{x}}_a = \widetilde{\mathbf{x}}_a - \widetilde{\mathbf{x}}_b \, ,
\end{equation}
where $\widetilde{\mathbf{x}}_b = \mathcal{G}_{NN} \left(\mathcal{H}(\mathbf{x}_b)\right)$. Reconstructed analysis increments, their evolution over recent decades and their comparison to reanalyses is shown in Supplementary~Information~Figs.~10--12.

\subsection*{Quantification of Hadley Circulation Strength}

The Hadley circulation strength is typically quantified using the mass-weighted zonal-mean stream function $\psi$ in the latitude-pressure ($\varphi$-$p$) plane\textsuperscript{\cite{Pikovnik2022}}. The stream function $\psi$ is derived from the zonally-averaged mass conservation equation by vertical integration of the zonal-mean meridional wind $[v]$\textsuperscript{\cite{Oort1996}} :
\begin{equation}\label{eq:hc_strength}
\psi(\varphi,p) = \frac{2 \pi R \cos\varphi}{g} \int_0^p [v](\varphi,p') \,\mathrm{d}p'
\end{equation}
where $R$ is Earth's radius, $g$ is gravity, $\varphi$ is latitude, and $p$ is pressure. Common HC strength metrics are based on single-point values (maxima or minima) of $\psi$ at a certain level or within the entire Hadley cell\textsuperscript{\cite{Pikovnik2022}}.

By meridionally averaging the zonal-mean stream function across the Northern Hadley cell (denoted by operator $\left<\cdot\right>$), the stream function can be approximated as:
\begin{equation}\label{eq:hc_here}
\tilde{\psi}(p) = \langle\psi(\varphi,p)\rangle \approx \left(\frac{2 \pi R \,\langle\cos{\varphi}\rangle}{g}\right) \int_0^p \left<[v]\right>(p')\, \mathrm{d}p'
\end{equation}
Supplementary~Information~Fig.~13 confirms that the error introduced by this approximation is negligible. The mean Hadley cell strength is then evaluated as\textsuperscript{\cite{Nguyen2013a}}: 
\begin{equation}\label{eq:hc_value}
\overline{\tilde{\psi}} = \frac{\sum_{k=1}^K \tilde{\psi}_k \Delta p_k}{\sum_{k=1}^K \Delta p_k}
\end{equation}
where $N$ is the number of pressure levels $p_k, k=1,\ldots,K$, with $p_0$ corresponding to 70~hPa and $p_K$ to 1000~hPa. Weights are $\Delta p_k = (p_{k+1}-p_{k-1})/2$, and $\tilde{\psi}_k=\tilde{\psi}(p_k)$.

For cross-verification of our results, we also evaluate NHC strength trend using the average stream-function metric (Extended~Data~Fig.~\ref{extfig:7}b,d,f,h). Note that the averaging across the stream-function field produces smaller magnitudes, in line with previous studies\textsuperscript{\cite{Pikovnik2022}}.

\subsection*{Physical Consistency Constraint}

As the MAE--GNN reconstructs the atmospheric state level-by-level, vertical coherence and mass conservation are not necessarily satisfied. To ensure physical consistency of the reconstructed Hadley circulation, we generate an ensemble of 150 MAE--GNN versions and apply filtering based on two criteria:
\begin{enumerate}
    \item \textbf{Meridional wind deviation}: This criterion ensures that the reconstructed vertical profile of mean meridional wind, $\overline{v}_k = \overline{v}(p_k)$, within the NHC remains within physically plausible limits. These limits are defined by the maximum relative spread across existing reanalyses over their record period. The relative spread between any two reanalyses ($i,j$) is defined as
    \begin{equation}
        \varepsilon^{i,j}  = \max_{i,j\neq i} \left(\frac{\sum_{k=1}^K \left|\overline{v}_k^i - \overline{v}_k^j\right|}{\sum_{k=1}^K \left|\overline{v}_k^i\right|}\right) \,
    \end{equation}
    where $k$ denotes the pressure levels $p_k$ from 50~hPa to 1000~hPa. Evaluating all permutations of the five reanalyses (20 pairs) yields a conservative upper bound of $\varepsilon=\max_{i,j}\varepsilon^{i,j}=0.44$. A MAE--GNN version is then discarded if its cumulative reconstruction deviation from the target reanalysis exceeds this threshold over the study period (e.g., 1980--2024 for ERA5):
    \begin{equation}
        \frac{\sum_{k=1}^K \left|\widetilde{\overline{v}}_k - \overline{v}_k\right|}{\sum_{k=1}^K \left|\overline{v}_k\right|} > \varepsilon \, .
    \end{equation}
    This constraint ensures that the GNN mapping from sparse observations back to the grid remains consistent with the structural uncertainty already present across reanalysis products.
    \newline
    
    \item \textbf{Surface mass continuity closure}: This criterion evaluates the closure of the Northern Hadley cell by evaluating the stream function at the nominal surface level ($p=1000$~hPa). Theoretically, for a perfectly closed circulation, $\tilde{\psi}$(1000~hPa) $=0$. In practice, reanalyses exhibit near-zero values when $\tilde{\psi}$ is evaluated at the bottom model level, but not at the bottom pressure level, which intersects the orography. Furthermore, only 13 pressure levels are used to evaluate $\tilde{\psi}$, introducing additional discretisation error. Among the reanalyses, MERRA-2 shows the largest absolute deviation, approximately $0.073\times 10^{11}$ kg s$^{-1}$ (Extended~Data~Fig.~\ref{extfig:5}a). We therefore apply a conservative tolerance and retain only those MAE--GNN versions, for which $|\tilde{\psi}(1000\;\textrm{hPa})|\leq 0.146\times 10^{11}$ kg s$^{-1}$, corresponding to twice the maximum deviation observed in standard reanalysis products.
\end{enumerate}
An MAE--GNN version is considered sufficiently accurate only when all six reconstructions (derived from five reanalyses and the radiosonde observations) satisfy both criteria. On average, this filtering retains approximately 45 of the 150 MAE--GNN versions. Examples of retained and discarded versions are shown in Supplementary~Information~Fig.~14.

\subsection*{Sensitivity to observational sampling}

To assess the sensitivity of the derived trends to the spatial sampling of the radiosonde network, we performed observation-denial experiments in which a controlled fraction $f$ (10\%, 20\% or 40\%) of radiosonde stations (or their reanalysis-equivalent locations) was removed prior to reconstruction. The removal was applied at the station level and held fixed over the full 1980--2024 period, preserving the large-scale structure of the observing network and avoiding artificial temporal variability.

To account for the strong spatial inhomogeneity of the radiosonde network, stations were removed probabilistically using weights based on their spatial representativeness. Specifically, each station was assigned a weight corresponding to the time-averaged spherical area it represents at a reference pressure level (500 hPa). These areas were estimated by constructing a spherical Voronoi tessellation at each of the 1104 monthly time stamps (46 years $\times$ 12 months $ \times$ 2 instances per month) and computing the area of the Voronoi cell associated with each observed station (Supplementary~Information~Fig.~15). The resulting cell areas were averaged over time, accounting for both the spatial extent represented by each station and the temporal availability of its data.
From the pool of stations observed at least once at the reference level, a fraction $f$ was sampled with probability proportional to the time-averaged Voronoi cell area. The selected stations were then consistently removed across all pressure levels and time steps by setting the corresponding nodes to masked (unobserved). This procedure ensures that stations in densely observed regions (e.g. Europe and North America) are less likely to be removed, whereas stations in sparsely observed regions (e.g. tropics and Southern Hemisphere, and in general over oceans) are more likely to be dropped out, thereby maintaining the large-scale density structure of the network under thinning (Supplementary~Information~Fig.~16).

The reconstruction procedure was subsequently applied as in the full-network case. An ensemble of realizations was generated, each with a different random station removal, and only those members satisfying the physical consistency criteria were retained for trend estimation.

\subsection*{Systematic difference and trend analysis}

\subsubsection*{Systematic difference estimation and significance testing}

Let us denote the radiosonde‑based reconstructions (MAE--OBS) as
\begin{equation}\label{eq:radiosonde_rec}
    \widetilde{\mathbf{x}}_y^i(t_j;p_k) = \mathcal{G}_{NN}^i \left(\mathbf{y}(t_j;p_k)\right)
\end{equation} 
and the reanalysis-sampled reconstructions as
\begin{equation}\label{eq:reanalysis_rec}
    \widetilde{\mathbf{x}}_a^i(t_j;p_k) = \mathcal{G}_{NN}^i \left(\mathcal{H}\left(\mathbf{x}_a(t_j;p_k)\right)\right) \, ,
\end{equation}
where each ensemble member $i=1,\ldots,N$ uses a distinct MAE mapping $\mathcal{G}_{NN}^i$, $t_j$, $j=1,\ldots,T$ represent the time instances and $k=1,\ldots,K$ the pressure levels. $\mathbf{y}(t_j; p_k)$ denotes radiosonde observations and $\mathbf{x}_a(t_j; p_k)$ denotes reanalysis fields, with $\mathcal{H}$ representing the observation operator that maps reanalysis fields to the observation locations.

All MAE--OBS reconstructions share the same observational input $\mathbf{y}(t_j; p_k)$, and all reanalysis-based reconstructions share the same reanalysis input sampled at the observation locations, $\mathcal{H}(\mathbf{x}_a(t_j; p_k))$. Differences across ensemble members of the same data type (radiosonde or reanalysis) therefore reflect uncertainty associated with the MAE mapping rather than independent data realisations.

For each ensemble member, annual-mean stream function (or mean meridional wind averaged over the Northern Hadley cell) time series were available for the reanalysis period (1 sample per year, typically 1980--2024, with shorter periods for JRA-55 and NOAA~20CRv3).

For the purpose of bias estimation, ensemble means of annual-mean MAE--OBS reconstructions (Eq.~\ref{eq:radiosonde_rec}), $\left< \widetilde{\mathbf{x}}_y\right>(t;p_k)$ and reanalysis-based reconstructions (Eq.~\ref{eq:reanalysis_rec}), $\left< \widetilde{\mathbf{x}}_a\right>(t;p_k)$, were first computed separately for every year and each pressure level. Thus, any systematic differences isolate the effect of the input data (radiosonde observations versus reanalysis equivalents) rather than differences in the reconstruction method. The paired \textit{t}-test was then applied to the resulting time-series of paired ensemble-mean values to assess whether the mean difference over the analysis period differs significantly from zero.

Mean differences shown in the figures correspond to the climatological (time-mean) difference of the ensemble-mean reanalysis-based and radiosonde-based reconstructions. Uncertainty ranges represent 95\% confidence intervals derived from the paired \textit{t}-test, computed as the product of the critical \textit{t} value (evaluated at the 0.975 quantile) and the standard error of the ensemble-mean difference. The number of degrees of freedom is equal to the length of the analysed time period minus one. Differences whose confidence intervals do not include zero are considered statistically significant ($p<0.05$).

\subsubsection*{Trend significance in an ensemble of time series}

For each retained reconstruction $i = 1, \ldots, N$, linear trends in Northern Hadley cell (NHC) strength were estimated from the annual‑mean stream function time series $\psi_i(t)$ (where $t$ denotes time in years) using ordinary least‑squares regression,
\begin{equation}
\psi_i(t) = \alpha_i + \beta_i\, t + \varepsilon_i(t),
\end{equation}
where $\beta_i$ denotes the linear trend (slope), $\alpha_i$ the constant, and $\varepsilon_i(t)$ the residual.
This regression yields, for each reconstruction, an individual trend estimate $\hat{\beta}_i$ and its associated standard error $\sigma_{\hat{\beta}_i}$.

To obtain a representative trend across reconstructions, individual trend estimates were combined using inverse‑variance weighting, which provides the statistically optimal estimator when uncertainties differ among reconstructions.
The weights are defined as
\begin{equation}
w_i = \frac{1}{\sigma_{\hat{\beta}_i}^2}.
\end{equation}
The combined trend estimate is then given by
\begin{equation}
\hat{\beta}_{\mathrm{comb}} =
\frac{\sum_{i=1}^{N} w_i\, \hat{\beta}_i}{\sum_{i=1}^{N} w_i},
\end{equation}
with an associated standard error obtained from the inverse of the summed weights
\begin{equation}
\sigma_{\hat{\beta}_{\mathrm{comb}}}
= \left( \sum_{i=1}^{N} w_i \right)^{-1/2}.
\end{equation}

A two‑sided 95\% confidence interval for the combined trend was constructed as
\begin{equation}
\hat{\beta}_{\mathrm{comb}} \;\pm\;
t_{0.975,\;\nu}\,\sigma_{\hat{\beta}_{\mathrm{comb}}},
\end{equation}
where $t_{0.975,\;\nu}$ denotes the 97.5th percentile of the Student‑$t$ distribution with
\begin{equation}
\nu = N - 2
\end{equation}
degrees of freedom, accounting for the estimation of both linear trend $\beta$ and constant $\alpha$.

The resulting confidence intervals quantify the sensitivity of the inferred trend to the MAE mapping, rather than sampling uncertainty of the underlying observations.

\subsubsection*{Trend significance in latitude--pressure space}

For analyses involving spatially resolved trends in latitude--pressure space, trend significance was assessed using the non-parametric Mann--Kendall test with trend-free pre-whitening, which accounts for serial correlation in the time series. This approach was applied where spatial maps of trends were analysed. 

Individual \textit{t} values and exact \textit{p} values for each reanalysis product are not explicitly reported. Instead, throughout the manuscript, statistical significance is indicated where the corresponding 95\% confidence interval excludes zero, equivalent to a two-sided test with $p<0.05$.


\providecommand{\backmatter}{}

\section*{Acknowledgements}
The authors are grateful to Lina Boljka, Massimo Bonavita, Tony McNally, Uroš Perkan, and Gregor Skok for their insightful comments and thorough review of the manuscript. The authors would like to acknowledge Bo\v{s}tjan Melinc for invaluable help with the composition and style of the figures. The authors would also like to thank Peter Bechtold, Carlo Buontempo, Sibo Cheng, Stephen English, Hans Hersbach, Michael Mayer, Paul Poli, Dinand Scheppers, Benoit Vanniere and Nedjeljka \v{Z}agar for fruitful discussions on the topic.

\section*{Declarations}

\subsection*{Funding}

Matic Pikovnik acknowledges funding from the Slovenian Research and Innovation Agency (ARIS) Programme P1-0188 and Grant 55351, and from the Slovenian National Recovery and Resilience Fund project ULTRA within subproject 6.03 Environmental Technologies for Climate Change Mitigation and Adaptation. \v{Z}iga Zaplotnik acknowledges the funding by the European Union under the Destination Earth initiative and Copernicus Climate Change Service (C3S). The authors acknowledge the financial support from the Slovenian Research and Innovation Agency (ARIS) through the UL VIP project (ARISE) under contract no. SN-ZRD/22-27/510.

\subsection*{Competing interests}

The authors declare no competing interests.

\subsection*{Ethics approval and consent to participate}

Not applicable.

\subsection*{Consent for publication}

Not applicable.

\subsection*{Data availability}

Radiosonde measurements data from the Integrated Global Radiosonde Archive Version 2.2 used for the analysis in this article are available at \url{https://www.ncei.noaa.gov/data/integrated-global-radiosonde-archive/archive/}. The ERA5 datasets are freely available from the open-access Climate Data Store (CDS) Catalogue, operated by the European Centre for Medium-Range Weather Forecasts (ECMWF) within the Copernicus Programme (\url{https://cds.climate.copernicus.eu/\#!/home}). The JRA-3Q reanalysis datasets are freely available from the National Centre for Atmospheric Research (NCAR) Data Archive web page (\url{https://gdex.ucar.edu/datasets/}). The JRA-55 reanalysis datasets are freely available from the National Centre for Atmospheric Research (NCAR) Data Archive web page (\url{https://gdex.ucar.edu/datasets/}). The MERRA-2 reanalysis datasets are freely available at the \url{https://disc.gsfc.nasa.gov/} web page, managed by the NASA Goddard Earth Sciences (GES) Data and Information Services Center (DISC). The NOAA 20CRv3 reanalysis datasets are freely available on the National Oceanic and Atmospheric Administration (NOAA) Physical Science Laboratory web page (\url{https://psl.noaa.gov/data/gridded/data.20thC\_ReanV3.html}).

\subsection*{Materials availability}

Not applicable.

\subsection*{Code availability}

The codes used for this study are available from the corresponding authors on reasonable request.

\subsection*{Author contribution}

Both authors contributed equally to this work. \v{Z}.Z. conceptualised the research and methodology, developed a novel masked autoencoder with a graph neural network, supervised the research and wrote the paper with input from M.P. M.P. obtained and processed the reanalyses and radiosonde data, developed the HC analysis codes, performed formal statistical analysis and visualisations.


\pagebreak
\section*{Extended Data}
\renewcommand{\figurename}{Extended~Data~Fig.}

\renewcommand{\thefigure}{\the\numexpr\value{figure}-5\relax}


\begin{figure}[h!]
\centering
\includegraphics[width=0.92\textwidth]{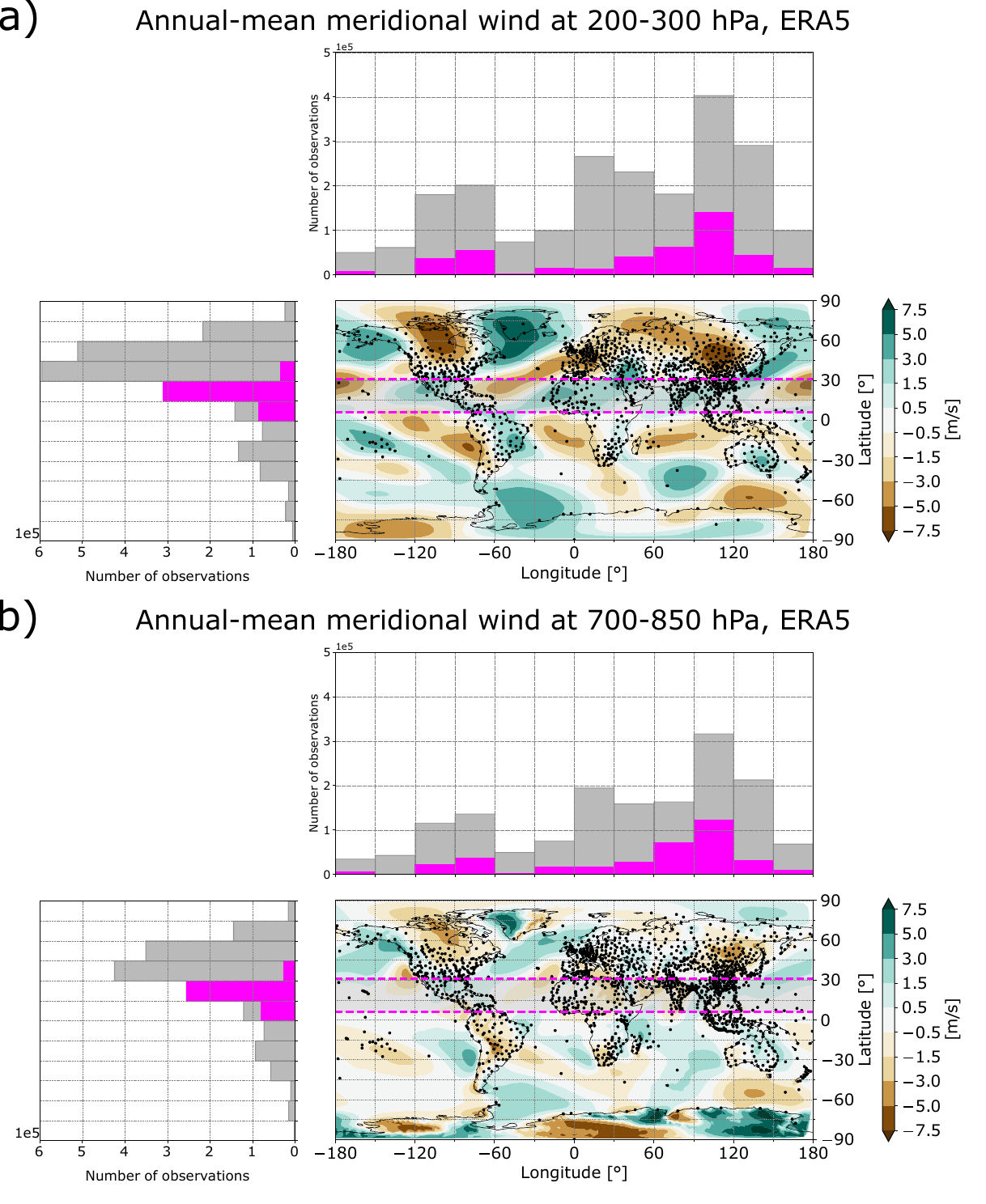}
\caption{\textbf{Spatial distribution of radiosonde observations.}
\textbf{a}, Climatological annual-mean meridional winds in the upper branch of the Hadley circulation (200--300~hPa), based on ERA5 data for 1980--2024. 
\textbf{b}, Same as \textbf{a}, but for the lower branch of the Hadley circulation (700--850~hPa).
In both panels, the region between the pink lines (6°N--31°N) denotes the climatological extent of the annual-mean Northern Hadley cell (NHC). Black dots indicate radiosonde observation locations at the corresponding pressure levels. Histograms along the top and left edges show the longitudinal and latitudinal distribution of radiosonde observations. Pink histograms indicate the corresponding distribution restricted to the NHC region.}
\label{extfig:1}
\end{figure}
\pagebreak

\clearpage
\begin{figure}[t!]
\centering
\includegraphics[width=0.98\textwidth]{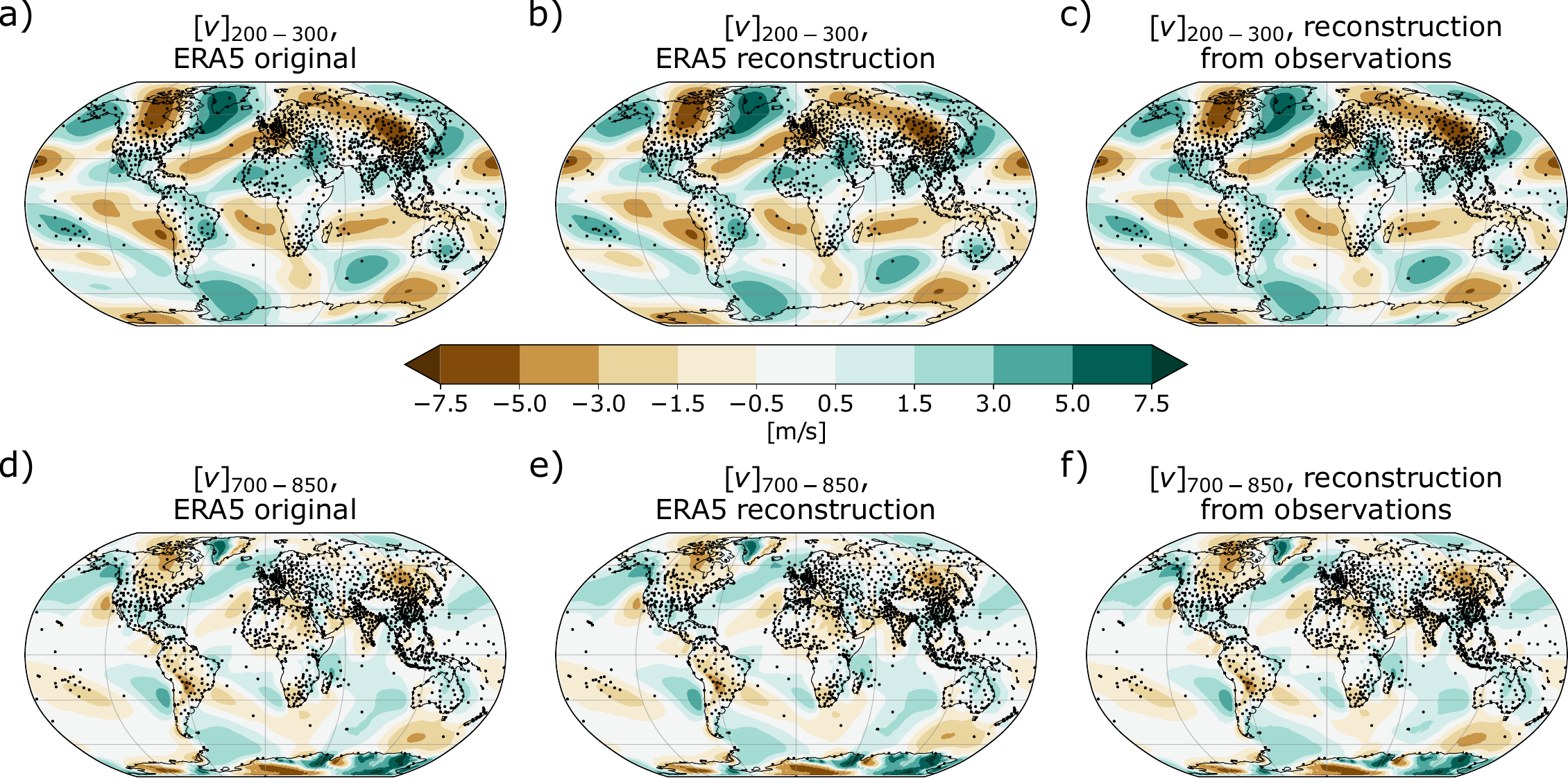}
\caption{\textbf{MAE--GNN reconstruction of global meridional winds from sparse observations.}
\textbf{a}, Climatological annual-mean meridional winds in ERA5 for the 200–300~hPa layer (1980--2024).
\textbf{b}, Reanalysis-based reconstruction using ERA5 values sampled at radiosonde locations.
\textbf{c}, Radiosonde-based reconstruction (MAE--OBS) derived directly from IGRAv2 radiosonde observations.
\textbf{d–f}, Same as \textbf{a–c}, but for the lower-tropospheric 700--850~hPa layer.
Black dots denote the radiosonde locations used for reconstruction.}
\label{extfig:2}
\end{figure}
\pagebreak

\clearpage
\begin{figure}[t!]
\centering
\includegraphics[width=0.98\textwidth]{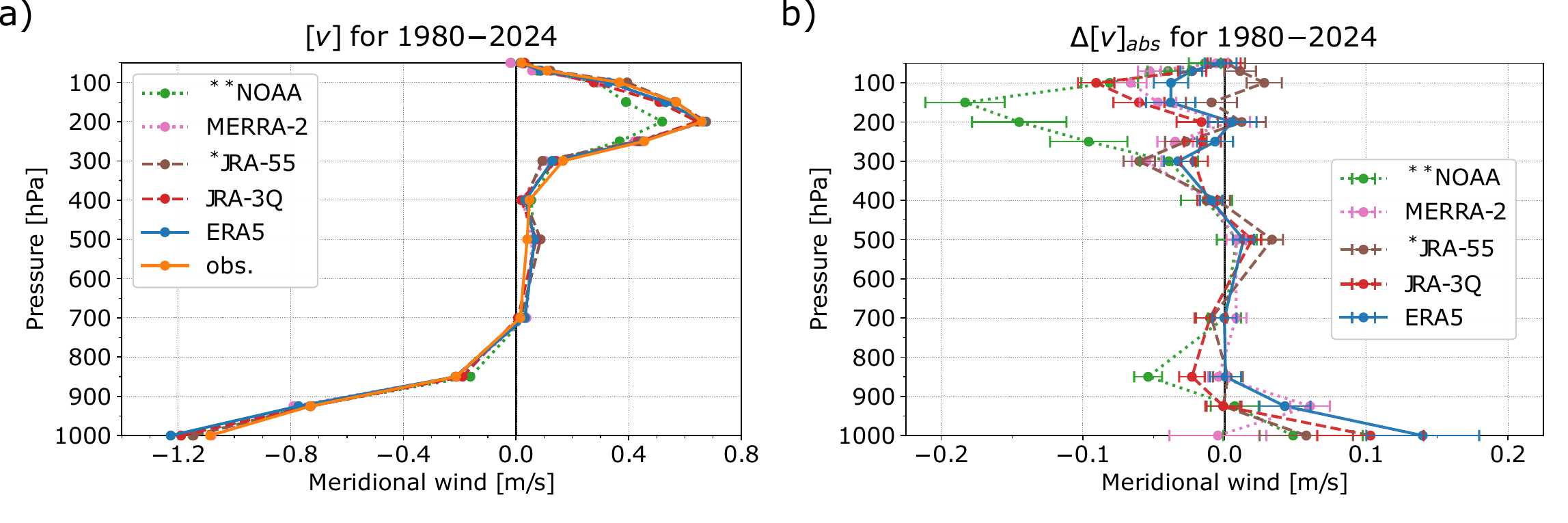}
\caption{\textbf{Meridional wind profiles within the Northern Hadley cell from radiosonde- and reanalysis-based reconstructions.}
\textbf{a}, Mean meridional wind profiles averaged over the Northern Hadley cell (6°N--31°N) for 1980--2024 ($^*$JRA-55 over 1980--2023; $^{**}$NOAA~20CRv3 over 1980--2015).
\textbf{b}, Differences between reanalysis-based and radiosonde-based reconstructions.
Error bars denote the 95\% confidence interval.}
\label{extfig:3}
\end{figure}
\pagebreak

\clearpage
\begin{figure}[t!]
\centering
\includegraphics[width=0.98\textwidth]{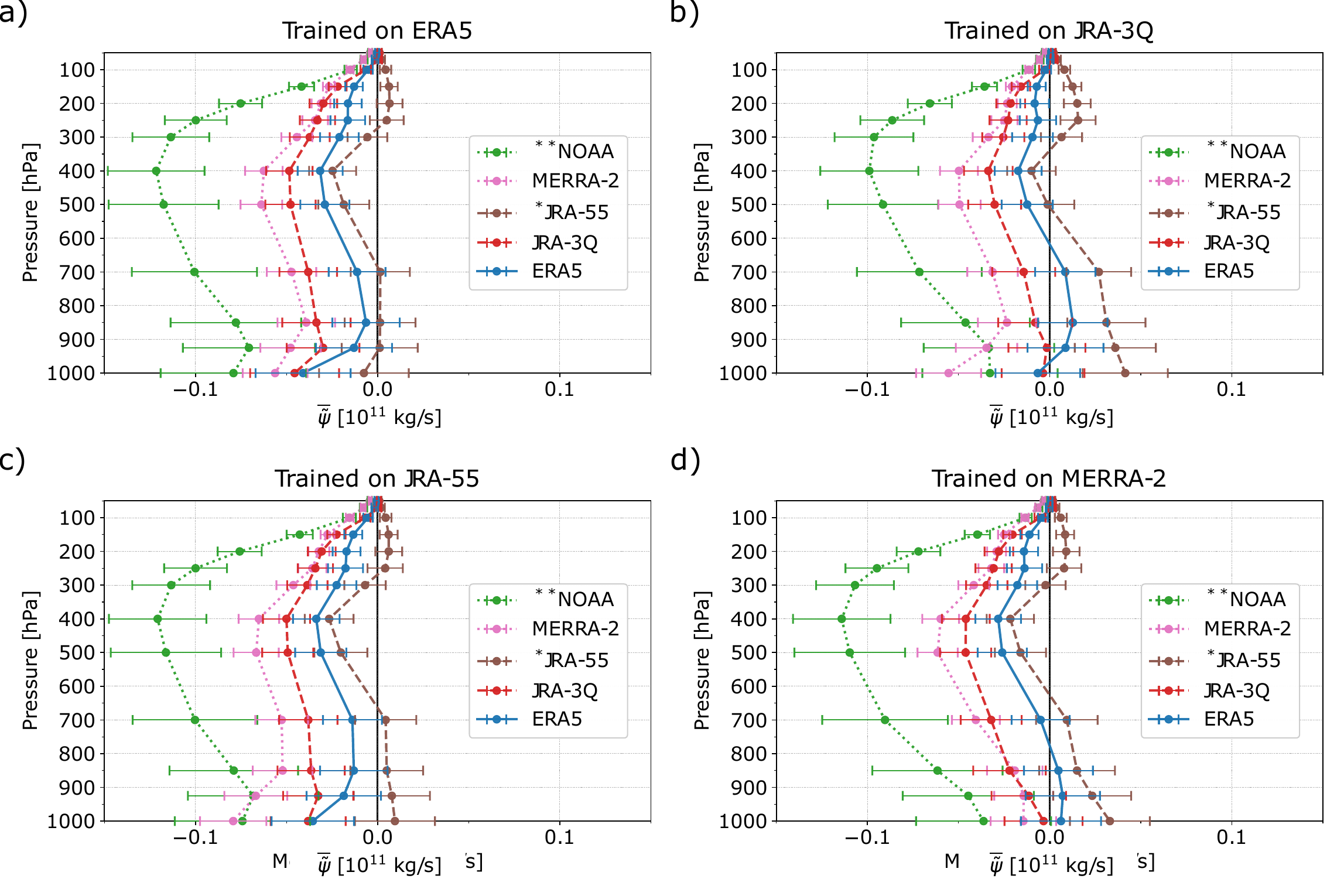}
\caption{\textbf{Differences in climatological NHC strength between reanalyses- and radiosonde-based reconstructions.}
Profiles show differences in stream-function strength derived from meridional wind reconstructions using different MAE models. Panels correspond to MAE--GNN models trained on:
\textbf{a}, ERA5 (1979--2024);
\textbf{b}, JRA‑3Q (1979--2024);
\textbf{c}, JRA‑55 (1979--2023);
\textbf{d}, MERRA‑2 (1980--2024). Negative values between 400 and 700 hPa (the level of peak NHC strength) indicate systematically stronger circulation in the radiosonde-based reconstructions. Error bars denote the 95\% confidence interval.}
\label{extfig:4}
\end{figure}
\pagebreak

\clearpage
\begin{figure}[t!]
\centering
\includegraphics[width=0.98\textwidth]{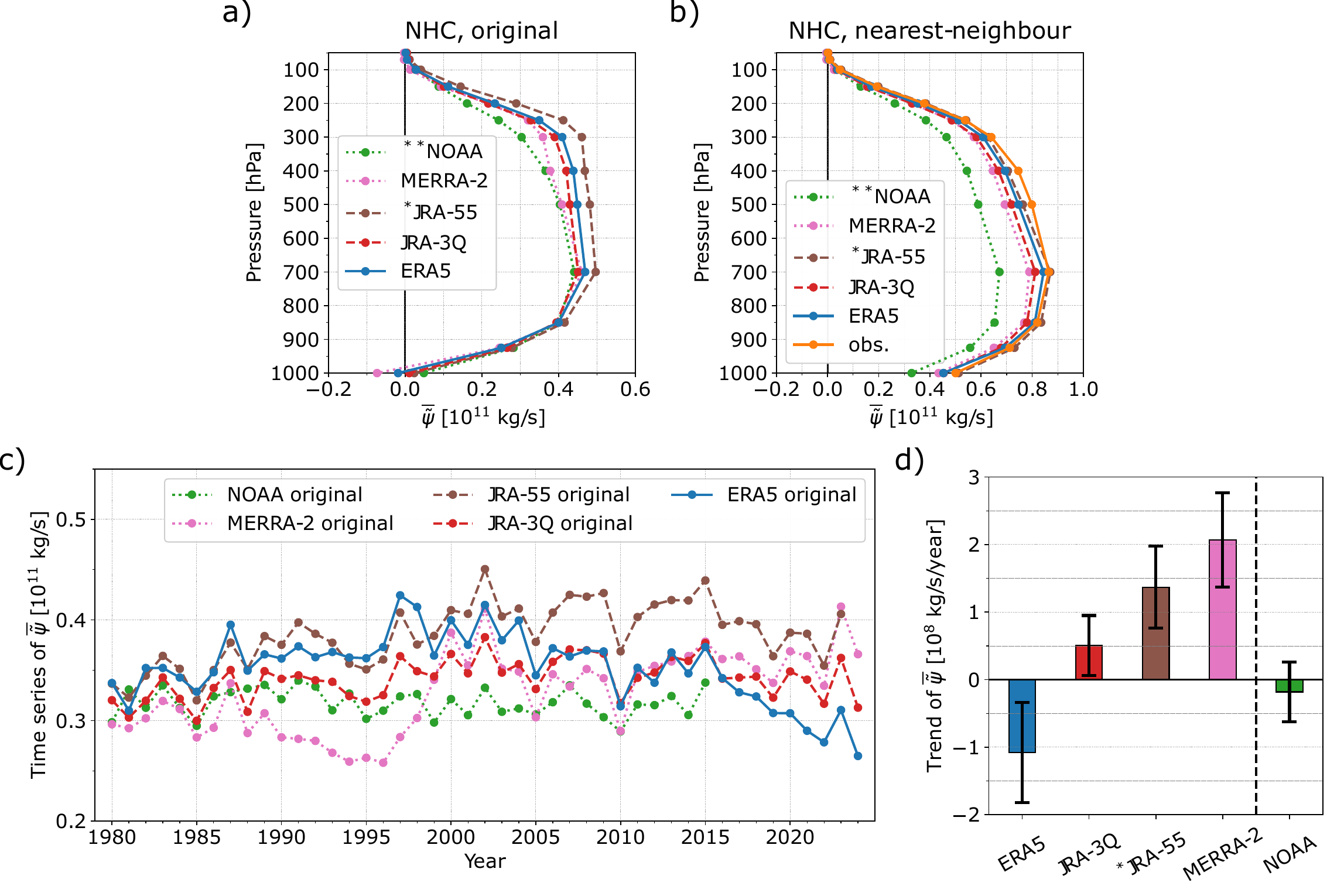}
\caption{\textbf{Climatological NHC stream function, time series, and trends in raw reanalyses.}
\textbf{a}, Stream-function profiles derived directly from raw reanalysis data.
\textbf{b}, Same as \textbf{a}, but computed from spherical nearest-neighbour interpolation of radiosonde data or their reanalysis equivalents in observation space.
\textbf{c}, Time series of the annual-mean stream function $\overline{\tilde{\psi}}$.
\textbf{d}, Linear trends in the annual-mean stream function $\overline{\tilde{\psi}}$.
Black bars denote the 95\% confidence intervals of the trend estimates. All panels are computed from raw reanalysis data over 1980--2024 for ERA5, JRA‑3Q and MERRA‑2, 1980--2023 for JRA‑55, and 1980--2015 for NOAA~20CRv3.}
\label{extfig:5}
\end{figure}
\pagebreak

\clearpage
\begin{figure}[t!]
\centering
\includegraphics[width=0.98\textwidth]{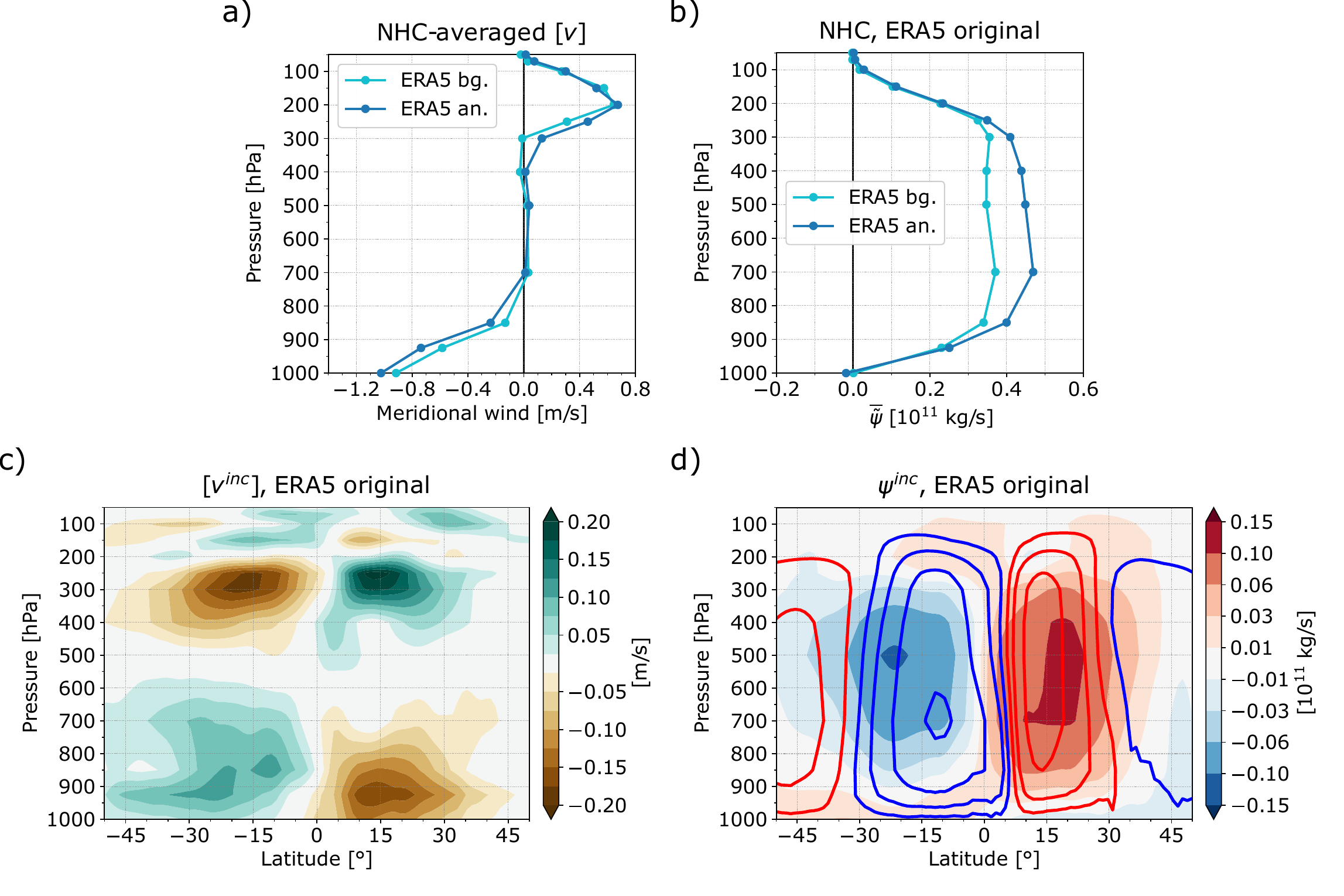}
\caption{\textbf{Effect of data assimilation on Hadley circulation strength in ERA5.}
\textbf{a}, Mean meridional wind profiles over the NHC (6°N--31°N) in ERA5 background and analysis.
\textbf{b}, Corresponding stream-function profiles.
\textbf{c}, Zonal-mean meridional wind analysis increments (analysis minus background).
\textbf{d}, Stream-function climatology (contours) and analysis increments (colours). Red contours denote positive stream-function values (0.1, 0.3, 0.6, $1\times 10^11$ kg s$^{-1}$) and blue their negative counterparts.
All panels show averages over 1980--2024.}
\label{extfig:6}
\end{figure}
\pagebreak

\clearpage
\begin{figure}[t!]
\centering
\includegraphics[width=0.63\textwidth]{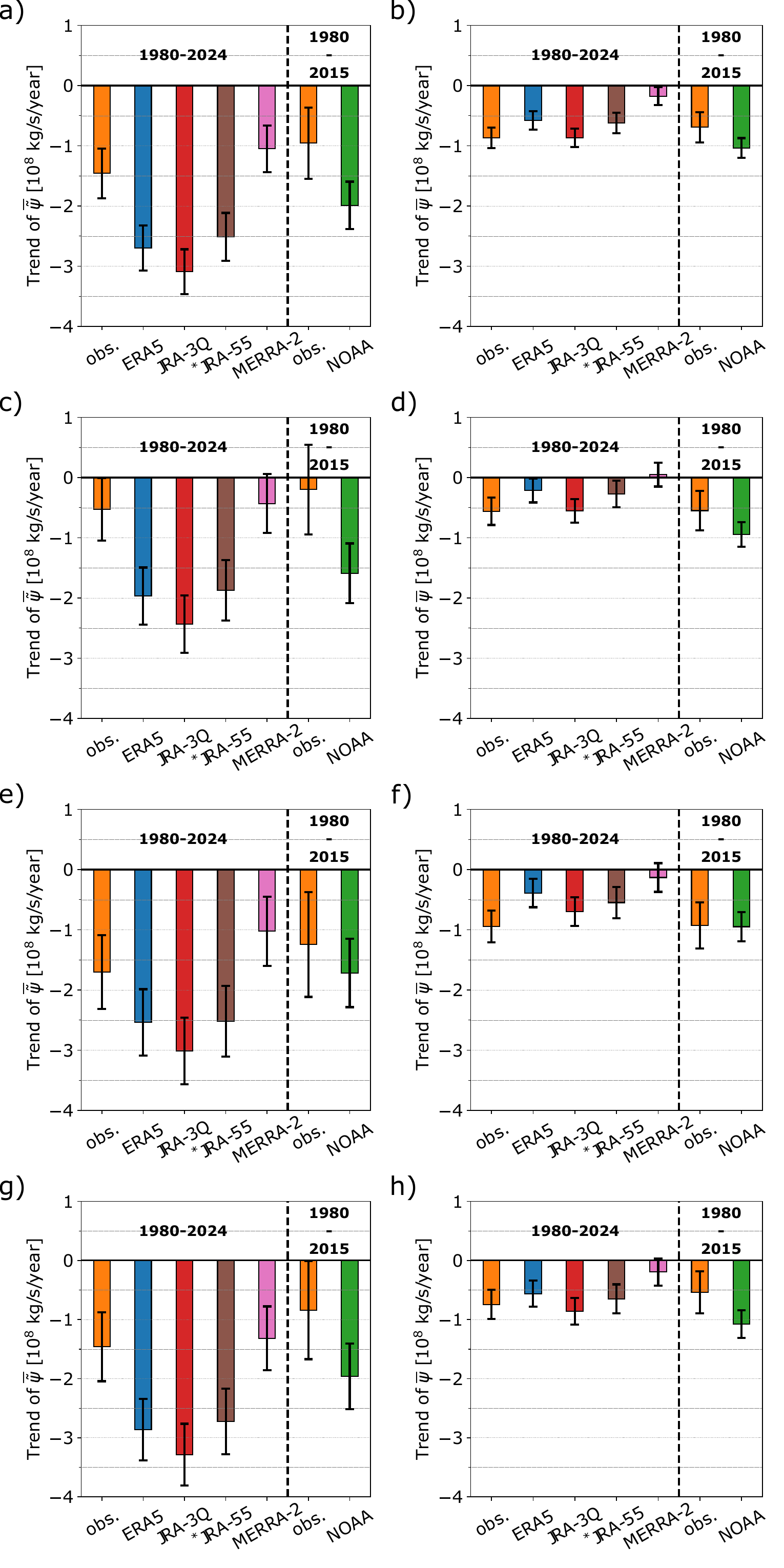}
\caption{\textbf{Sensitivity of NHC strength trends to training dataset and metric definition.}
Multidecadal trends in NHC strength in reconstructions for the 1980--2024 ($^*$JRA-55 for 1980--2023) and 1980--2015 periods. Rows indicate the reanalysis used for MAE training: \textbf{a,b}, ERA5; \textbf{c,d}, JRA-3Q; \textbf{e,f}, JRA-55; \textbf{g,h}, MERRA-2. Columns indicate the circulation metric: \textbf{a,c,e,g}, stream-function metric (Eqs.~\ref{eq:hc_here}, \ref{eq:hc_value}); \textbf{b,d,f,h}, average stream-function metric\textsuperscript{\cite{Pikovnik2022}}. Error bars denote the 95\% confidence intervals.}
\label{extfig:7}
\end{figure}
\pagebreak

\clearpage
\begin{figure}[t!]
\centering
\includegraphics[width=0.98\textwidth]{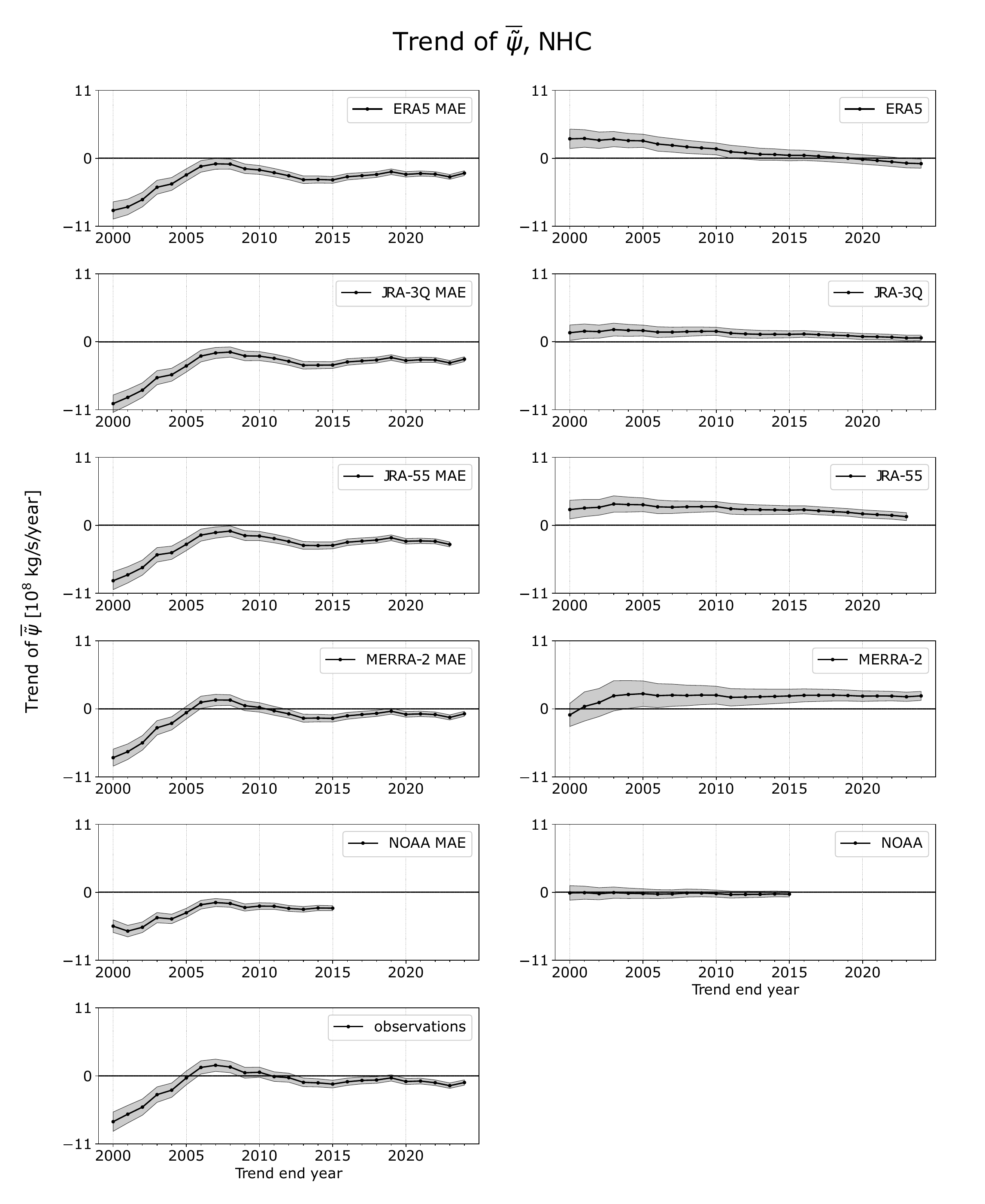}
\caption{\textbf{Dependence of NHC strength trends on analysis period length.}
Trends are computed for periods starting in 1980 and ending in the year indicated on the x-axis. Left panels show MAE--GNN reconstructions from different reanalyses and observations (MAE--OBS); right panels show raw reanalysis data. Shading denotes 95\% confidence intervals.}
\label{extfig:8}
\end{figure}
\pagebreak

\clearpage
\begin{figure}[t!]
\centering
\includegraphics[width=0.98\textwidth]{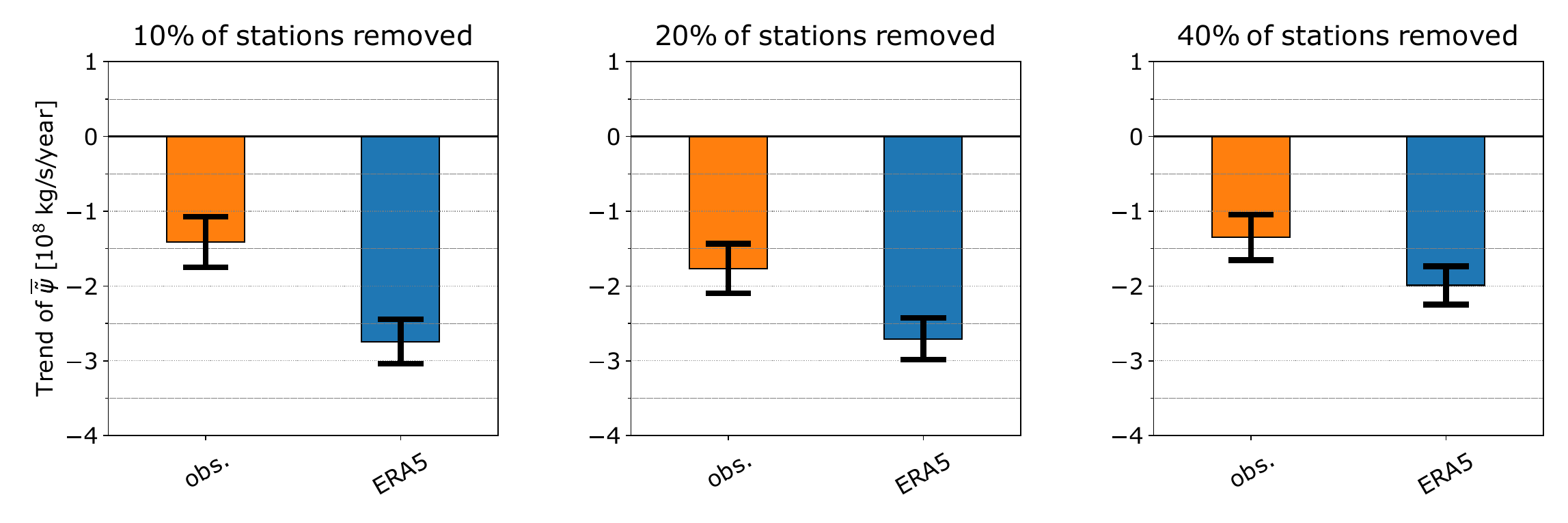}
\caption{\textbf{Sensitivity of NHC strength trends to observational sampling.}
Linear trends in the annual-mean stream function strength $\overline{\tilde{\psi}}$ for reanalysis-based reconstructions (ERA5) and radiosonde-based reconstructions (MAE--OBS), computed for three different levels of station removal (10\%, 20\% and 40\% of stations removed), demonstrating the robustness of the inferred trends to observational sampling. Black bars denote the 95\% confidence intervals of the trend estimates.}
\label{extfig:9}
\end{figure}
\pagebreak

\clearpage
\begin{figure}[t!]
\centering
\includegraphics[width=0.98\textwidth]{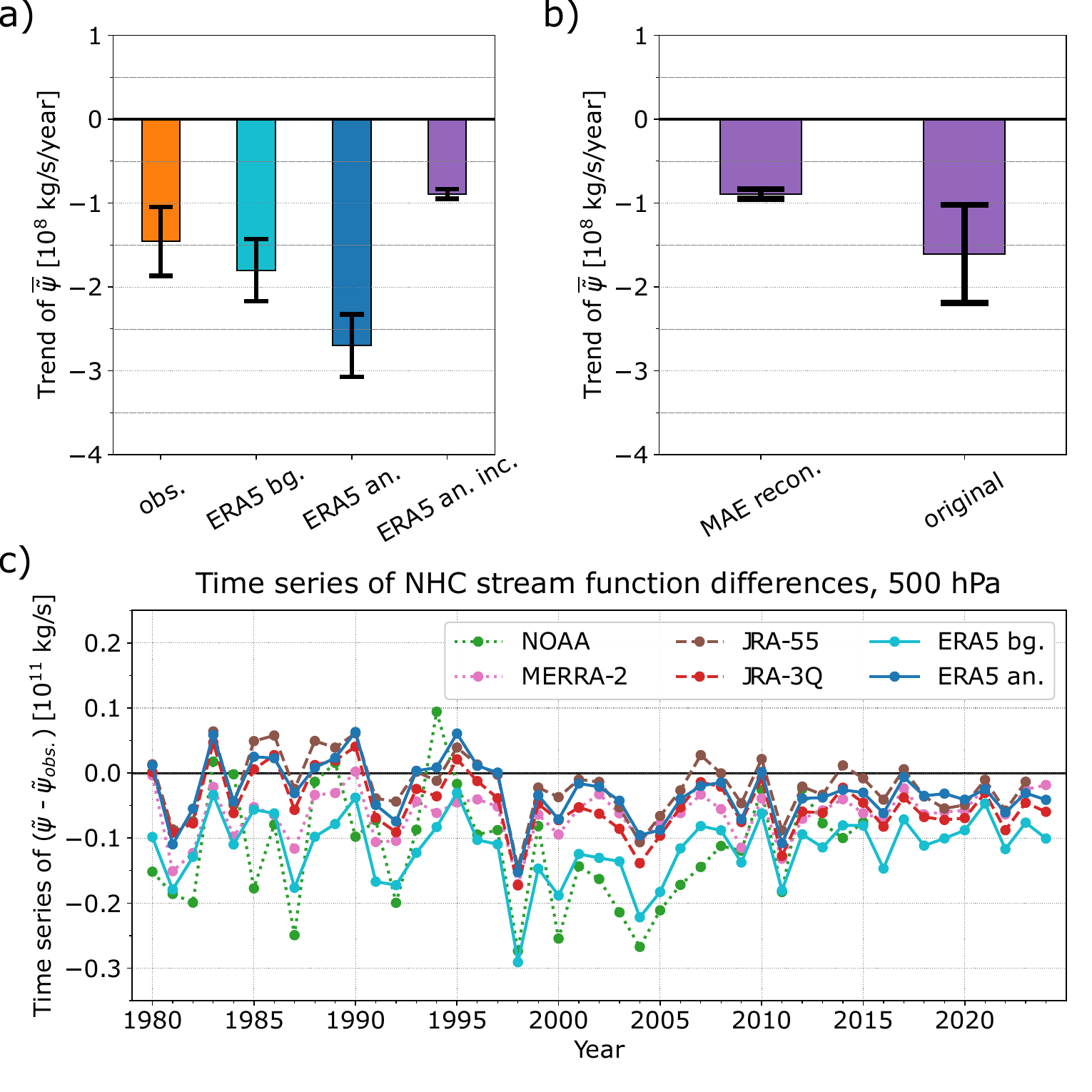}
\caption{\textbf{Temporal evolution of Northern Hadley cell strength.}
\textbf{a}, Trends in stream function from MAE reconstructions derived from observations, ERA5 background, ERA5 analysis and ERA5 analysis increments, respectively, over the 1980--2024 period. \textbf{b}, Trends in analysis increments from MAE reconstructions and ERA5 reanalysis. Error bars in \textbf{a},\textbf{b}, denote the 95\% confidence intervals for the linear trend estimates. \textbf{c}, Time series of differences in NHC stream function at 500 hPa (peak NHC strength) between radiosonde- and reanalysis-based reconstructions.}
\label{extfig:10}
\end{figure}


\end{document}


\maketitle

\renewcommand{\figurename}{Supplementary information Fig.}
\captionsetup[figure]{
    labelfont=bf
}

\begin{figure}[h!]
\centering
\includegraphics[width=0.88\textwidth]{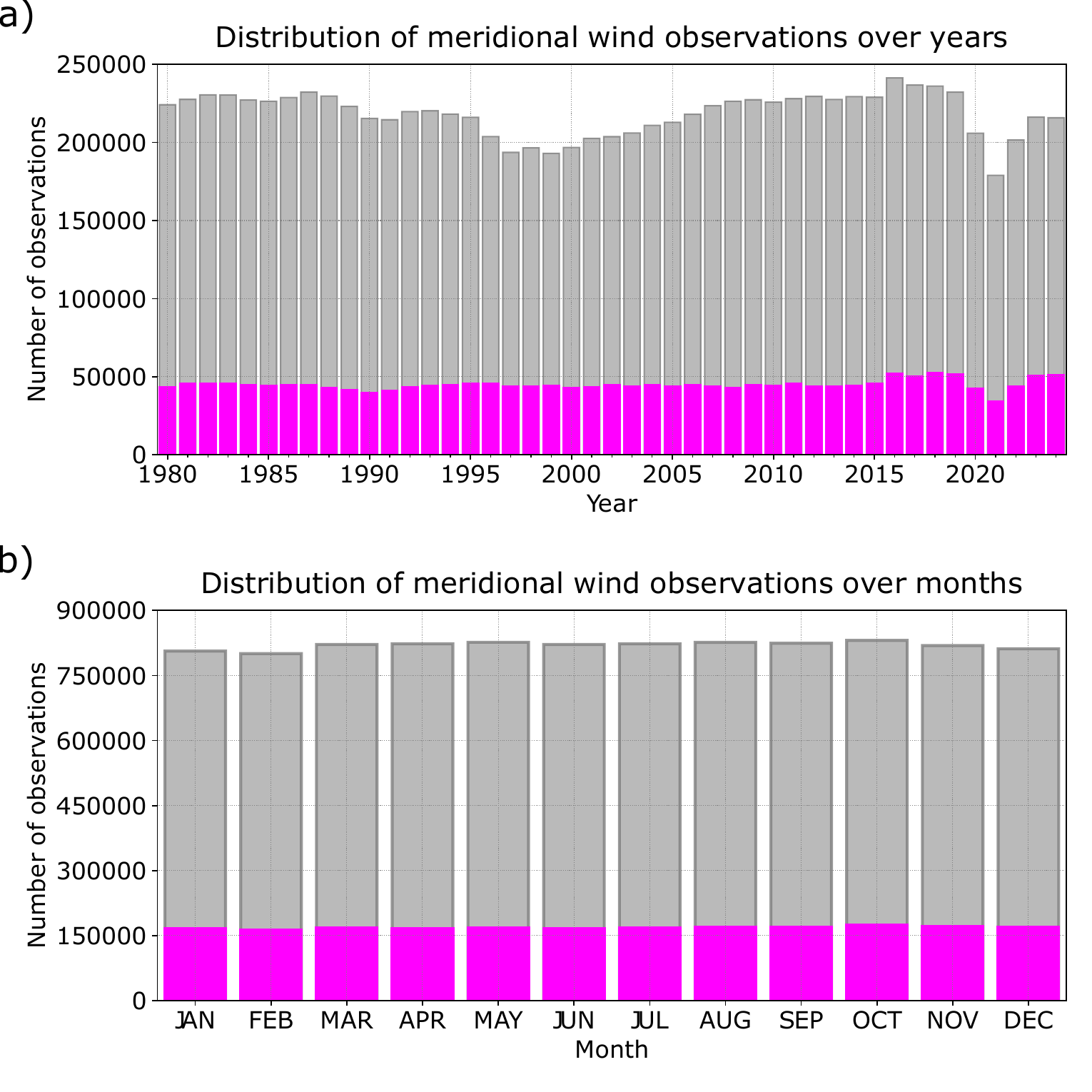}\setcounter{figure}{0}
\caption{\textbf{Temporal distribution of radiosonde meridional wind observations.} \textbf{a}, Annual distribution of meridional wind observations (1980–2024) in the IGRAv2 dataset. \textbf{b}, Monthly distribution. The reported numbers apply to all available (21) pressure levels. Grey bars show all observations; pink bars indicate observations within the NHC.}
\label{fig:fig_S01}
\end{figure}

\clearpage
\begin{figure}[t!]
\centering
\includegraphics[width=0.98\textwidth]{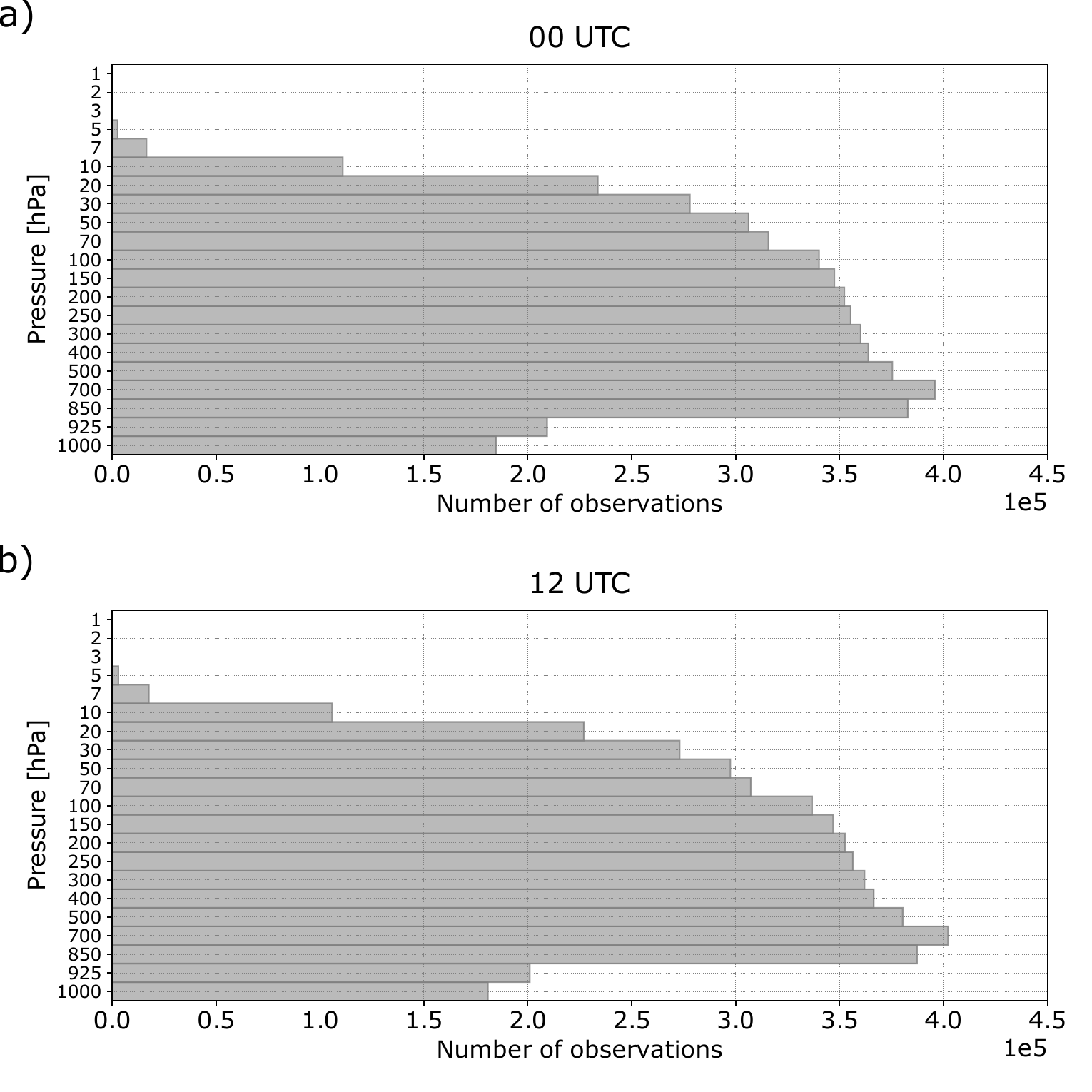}
\caption{\textbf{Vertical distribution of radiosonde meridional wind observations.} \textbf{a}, Observations at 00 UTC. \textbf{b}, Observations at 12 UTC. Results are shown for the IGRAv2 dataset (1980–2024).}
\label{fig:fig_S02}
\end{figure}
\pagebreak

\clearpage
\begin{figure}[t!]
\centering
\includegraphics[width=0.98\textwidth]{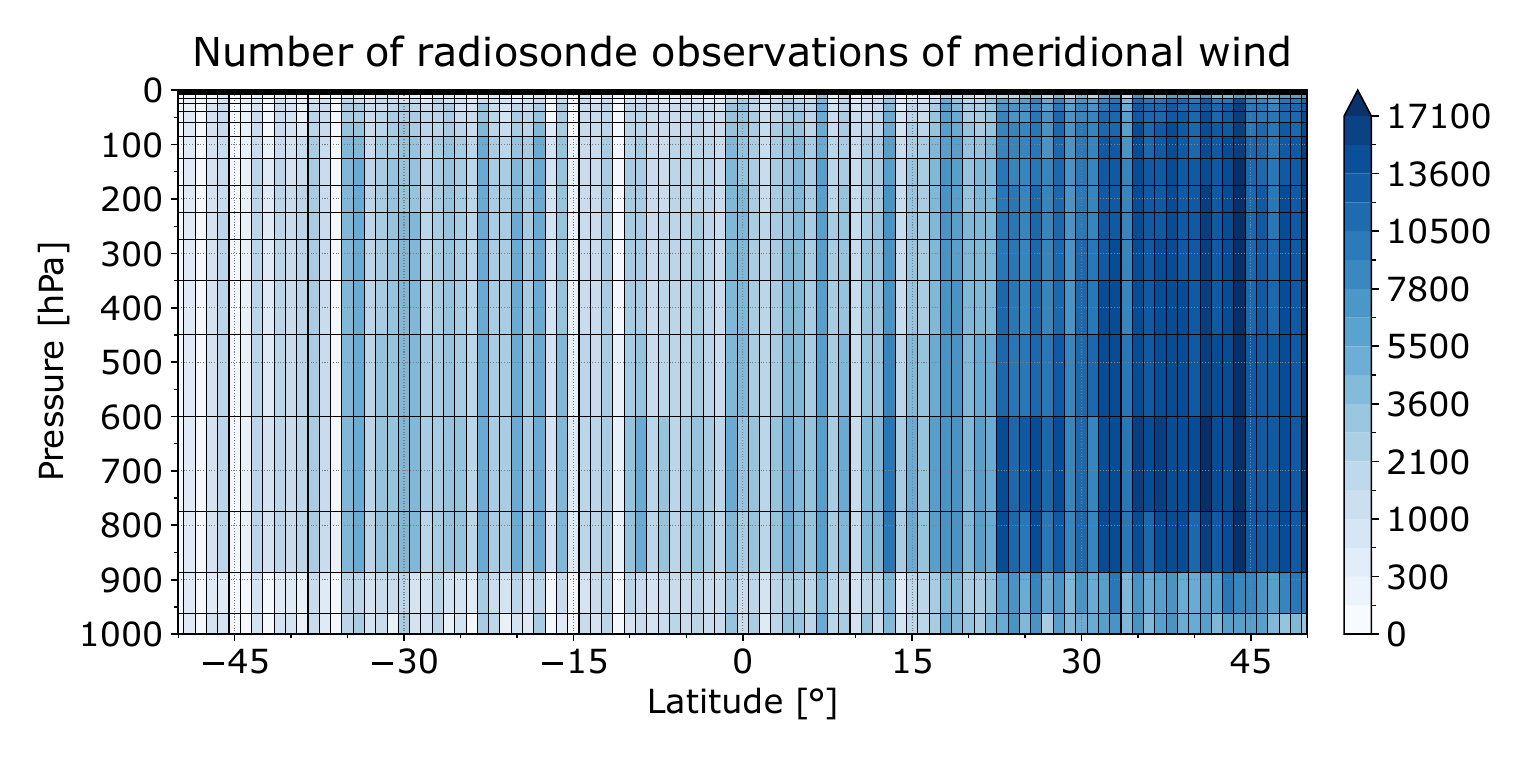}
\caption{\textbf{Vertical and meridional distribution of radiosonde meridional wind observations.} Density of meridional wind observations aggregated by latitude (1° bins) and pressure level over 1980–2024.}
\label{fig:fig_S03}
\end{figure}
\pagebreak

\clearpage
\begin{figure}[t!]
\centering
\includegraphics[width=0.98\textwidth]{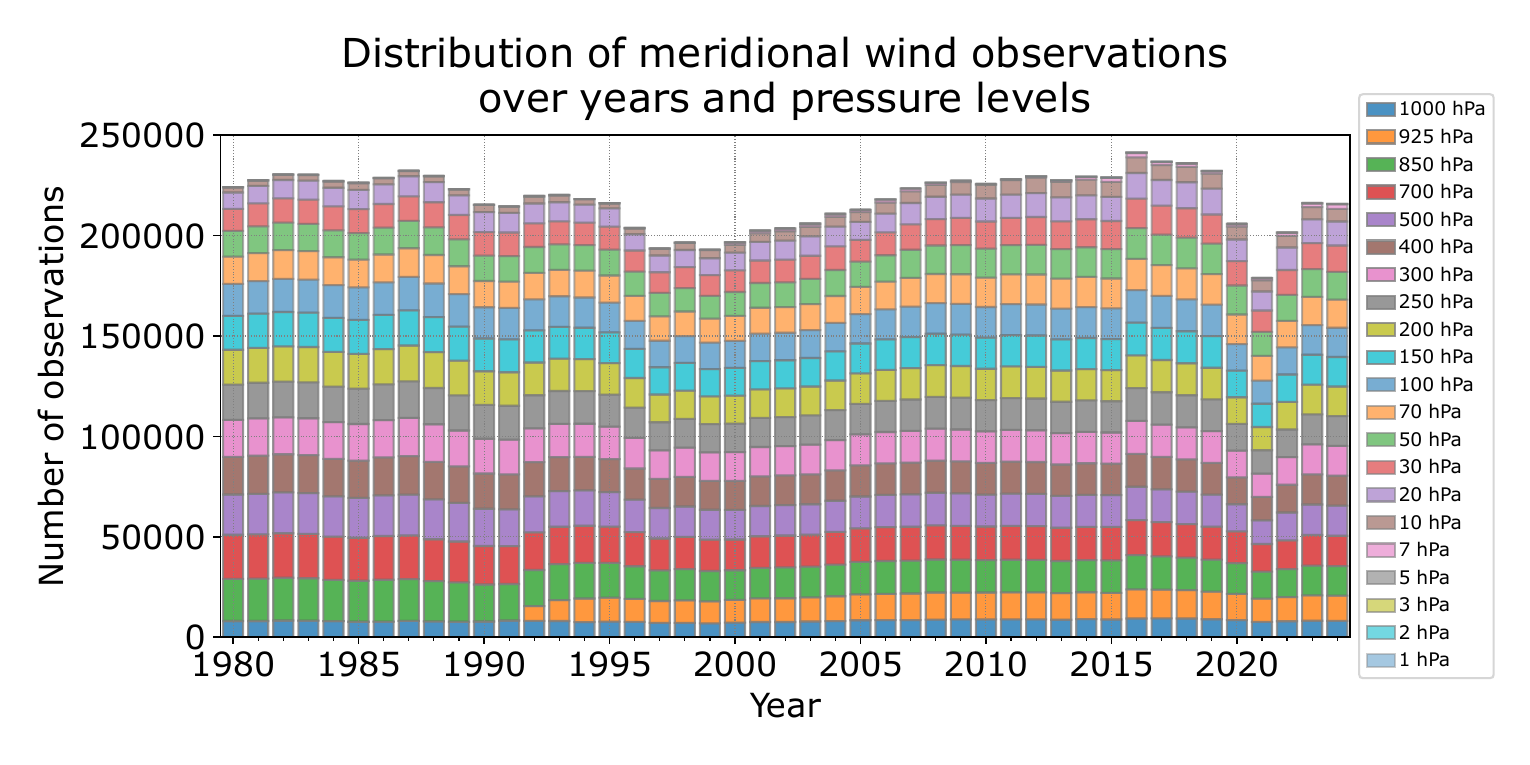}
\caption{\textbf{Temporal evolution of radiosonde sampling by pressure level.} Distribution of radiosonde observations over the years (1980–2024) and pressure levels (from 1 to 1000 hPa).}
\label{fig:fig_S04}
\end{figure}
\pagebreak

\clearpage
\begin{figure}[t!]
\centering
\includegraphics[width=0.98\textwidth]{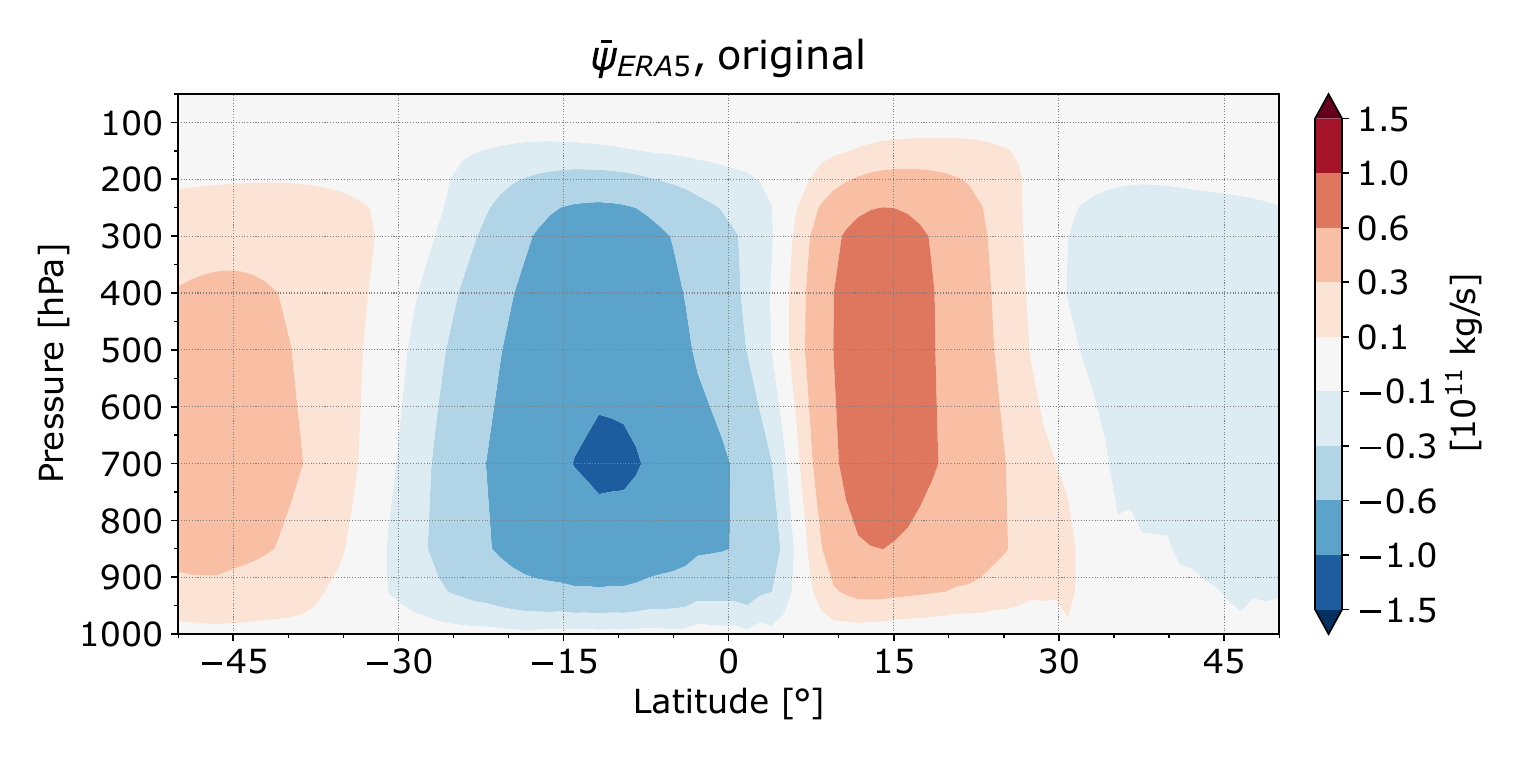}
\caption{\textbf{Stream-function representation of the climatological Hadley circulation.} Stream function of the annual-mean global Hadley circulation derived from ERA5 reanalysis data over 1980–2024.}
\label{fig:fig_S05}
\end{figure}
\pagebreak

\clearpage
\begin{figure}[t!]
\centering
\includegraphics[width=0.98\textwidth]{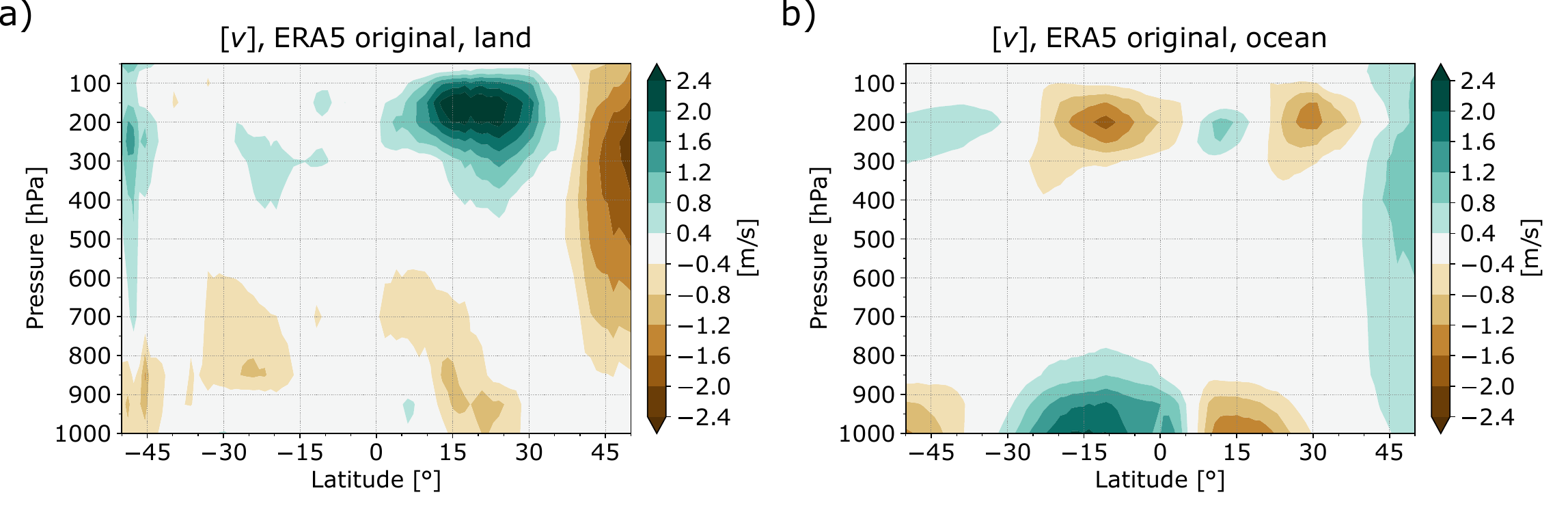}
\caption{\textbf{Influence of land–sea sampling on zonal-mean meridional wind.} Zonal-mean meridional wind, derived from ERA5 over 1980–2024 using: \textbf{a}, land-only columns; \textbf{b}, ocean-only columns.}
\label{fig:fig_S06}
\end{figure}
\pagebreak

\clearpage
\begin{figure}[t!]
\centering
\includegraphics[width=0.98\textwidth]{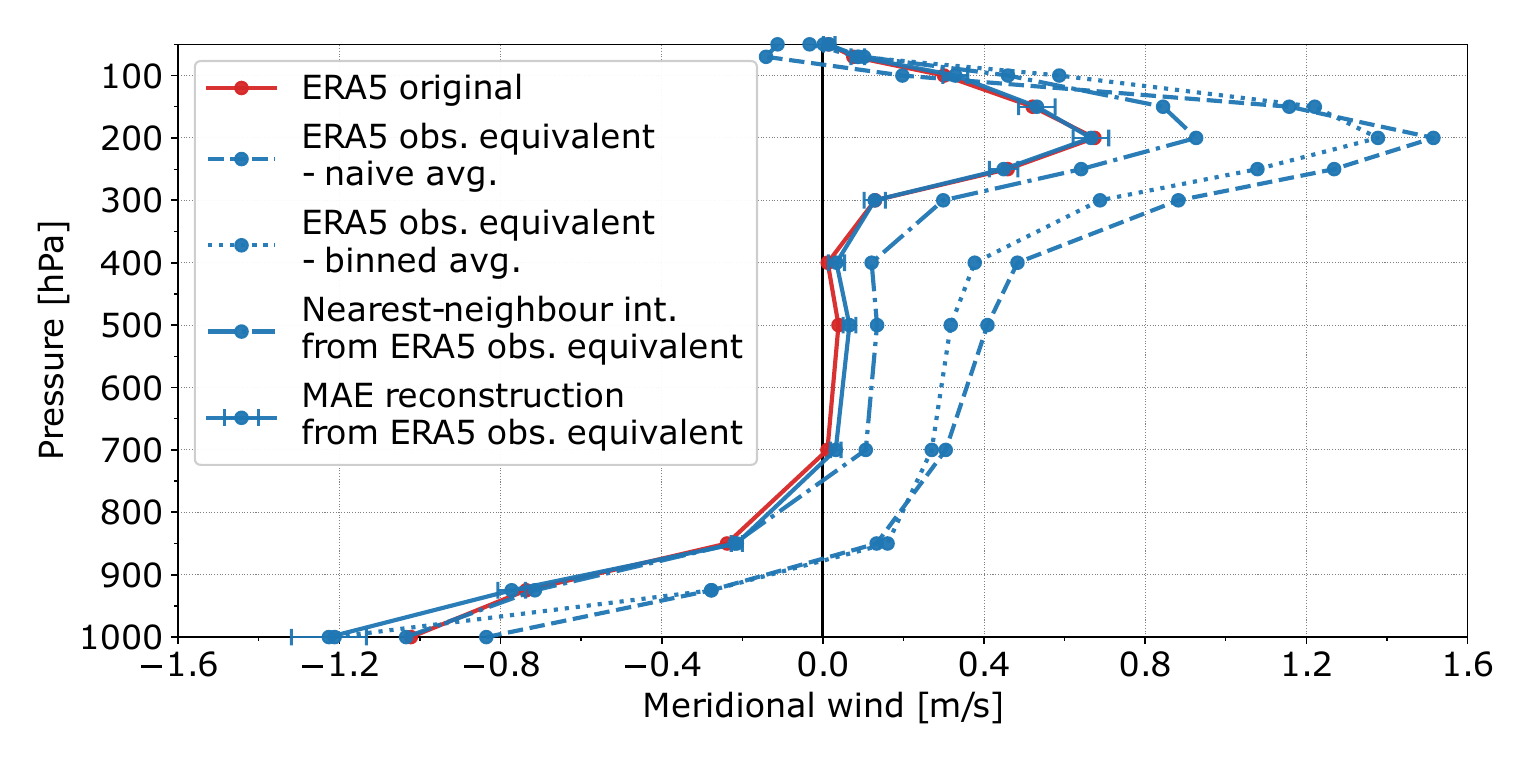}
\caption{\textbf{Comparison of averaging methods for NHC meridional winds.} Meridional winds in ERA5, averaged over 1980–2024 within the Northern Hadley cell (6°N–31°N). The averaging is performed on the raw ERA5 data (red) and in the observation space of the radiosondes using four different methods (blue): naive averaging (dashed), not accounting for data inhomogeneity; binned averaging (dotted), and NHC averaging after reconstructing the full space using nearest-neighbour interpolation from ERA5 equivalents of the observations (dashed dotted). Solid blue denotes NHC-averaged MAE reconstructions derived from ERA5 sampled at the observation locations. Error bars denote the standard deviation of the meridional wind among averaged reconstructions. The binning approach divides the latitude belt into six longitude bins, each 60° wide, computes the mean meridional wind within each bin, and then performs a zonal average.}
\label{fig:fig_S07}
\end{figure}
\pagebreak

\clearpage
\begin{figure}[t!]
\centering
\includegraphics[width=0.98\textwidth]{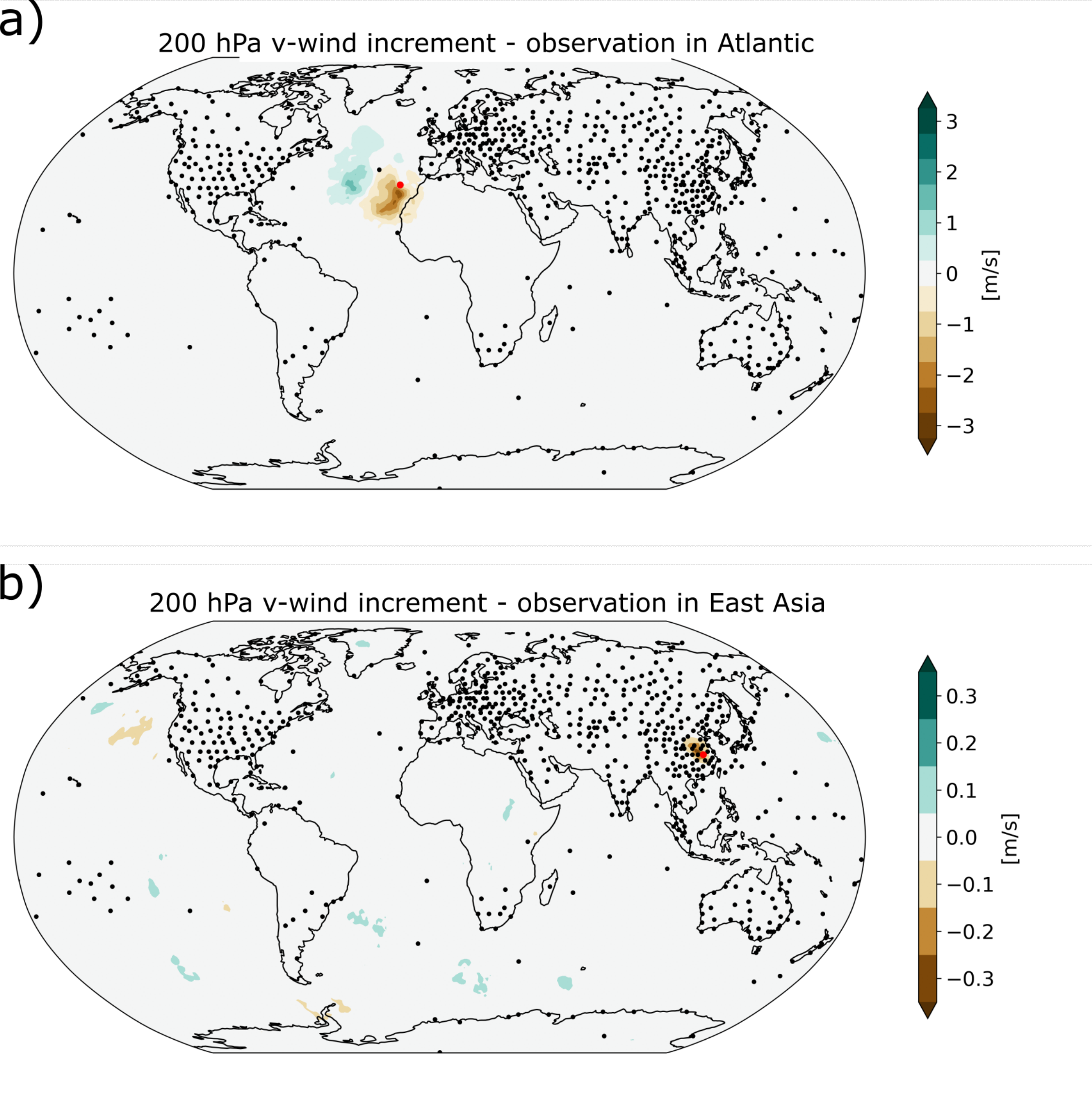}
\caption{\textbf{Spatial sensitivity of the MAE-GNN reconstruction to individual observations.} \textbf{a}, Impact on the reconstructed global meridional wind (increment) after adding a single observation in the data-sparse North Atlantic (red dot) to the existing radiosonde data. \textbf{b}, Corresponding impact after adding a single observation in the data-dense East Asia region. Note the different colour scales in \textbf{a} and \textbf{b}, indicating the stronger influence of individual observations in data-sparse regions.}
\label{fig:fig_S08}
\end{figure}
\pagebreak

\clearpage
\begin{figure}[t!]
\centering
\includegraphics[width=0.98\textwidth]{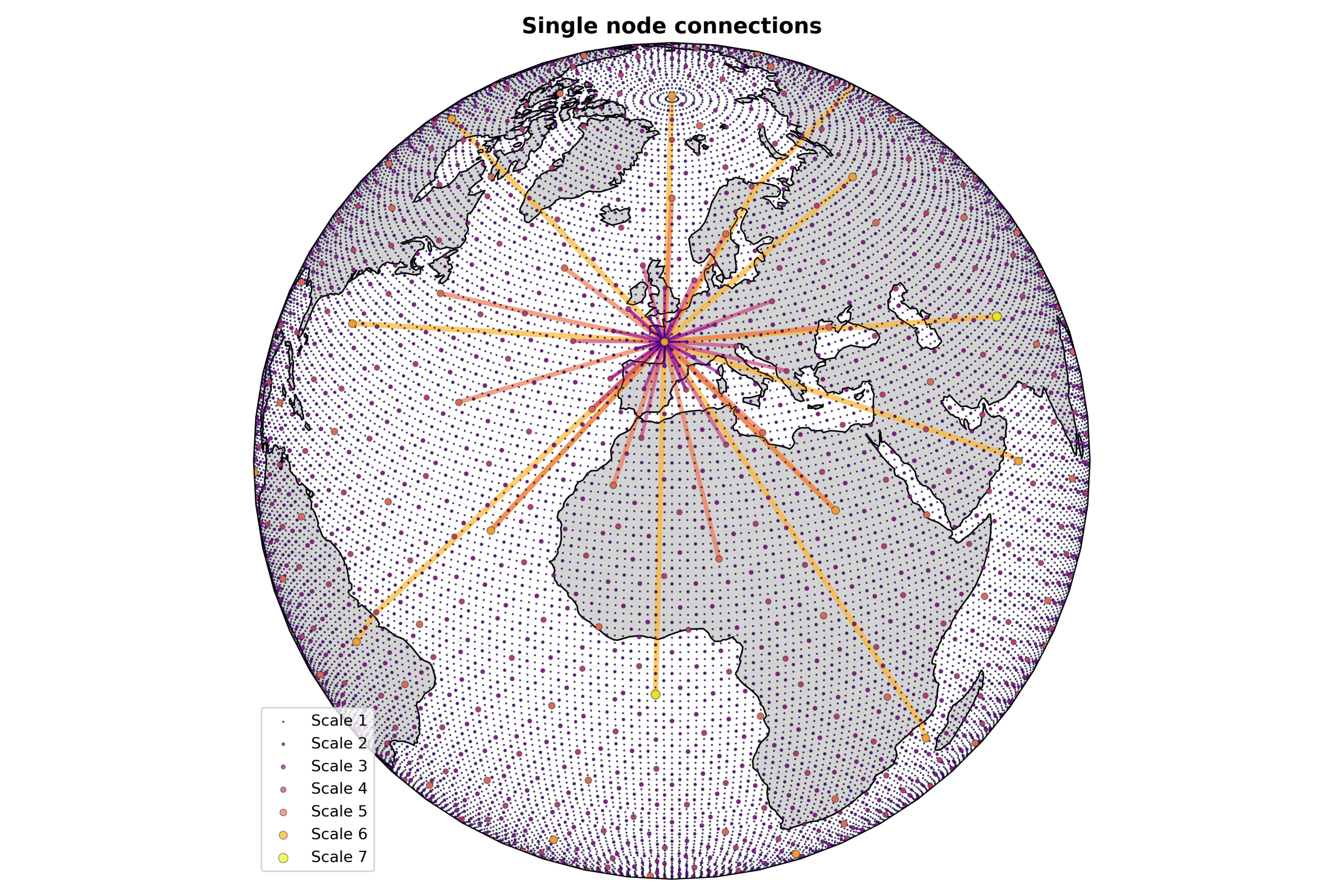}
\caption{\textbf{Multiscale graph connectivity in the MAE–GNN architecture.} Connections of a single scale-6 node in a multiscale graph. The multiscale graph consists of a total of 35 “scale 6” nodes.}
\label{fig:fig_S09}
\end{figure}
\pagebreak

\clearpage
\begin{figure}[t!]
\centering
\includegraphics[width=0.98\textwidth]{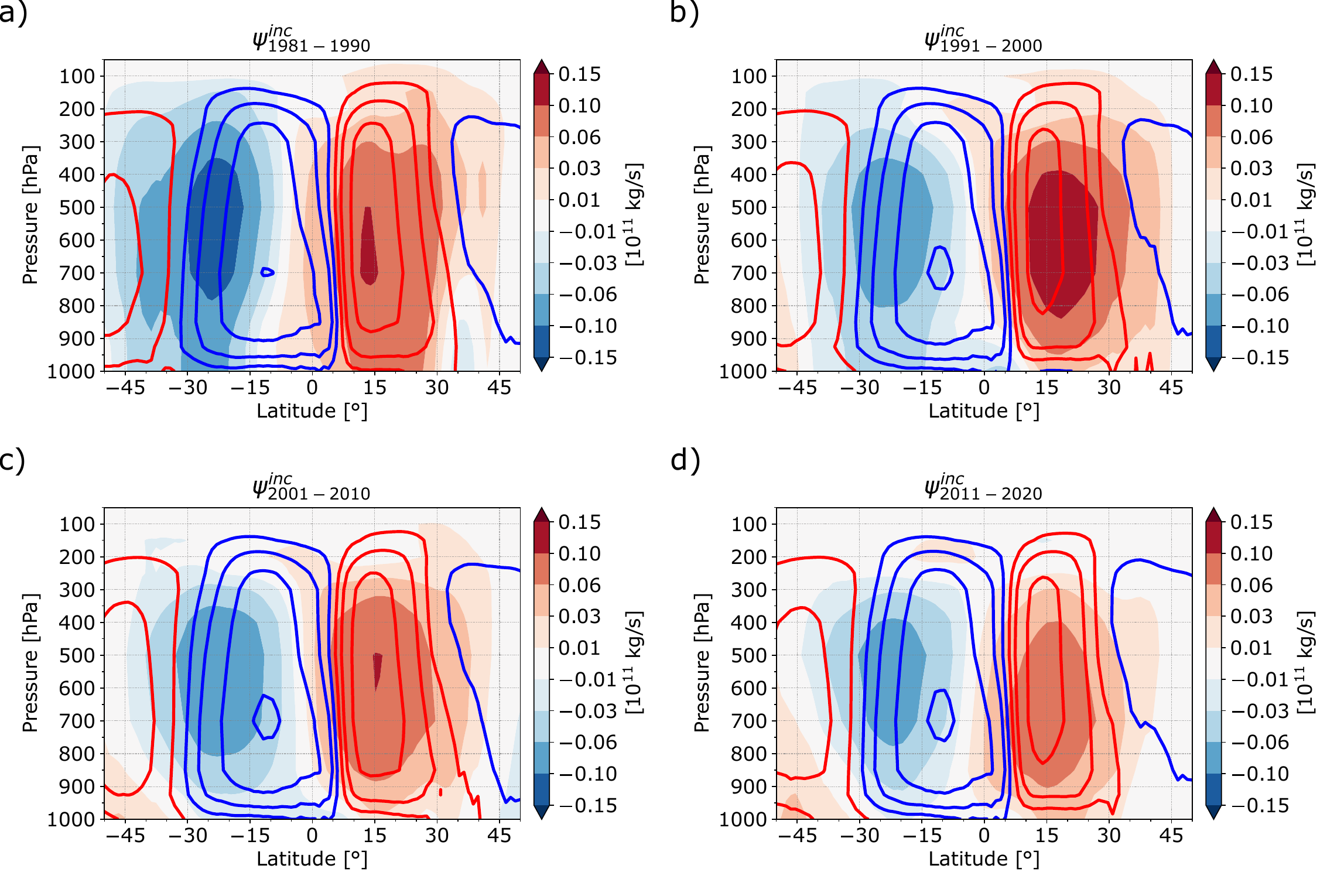}
\caption{\textbf{Decadal evolution of reconstructed Hadley circulation and ERA5 analysis increments.} Each panel displays the time-averaged Hadley circulation based on the MAE reconstructions derived from ERA5 analysis (contours; red contours for positive stream function values (0.1, 0.3, 0.6, 1 $\times$ 10$^{11}$ kg s$^{-1}$) and blue for negative counterparts) and the stream function of meridional wind analysis increments (colours). The panels show averages for: \textbf{a} 1981–1990; \textbf{b} 1991–2000; \textbf{c} 2001–2010; \textbf{d} 2011–2020.}
\label{fig:fig_S10}
\end{figure}
\pagebreak

\clearpage
\begin{figure}[t!]
\centering
\includegraphics[width=0.98\textwidth]{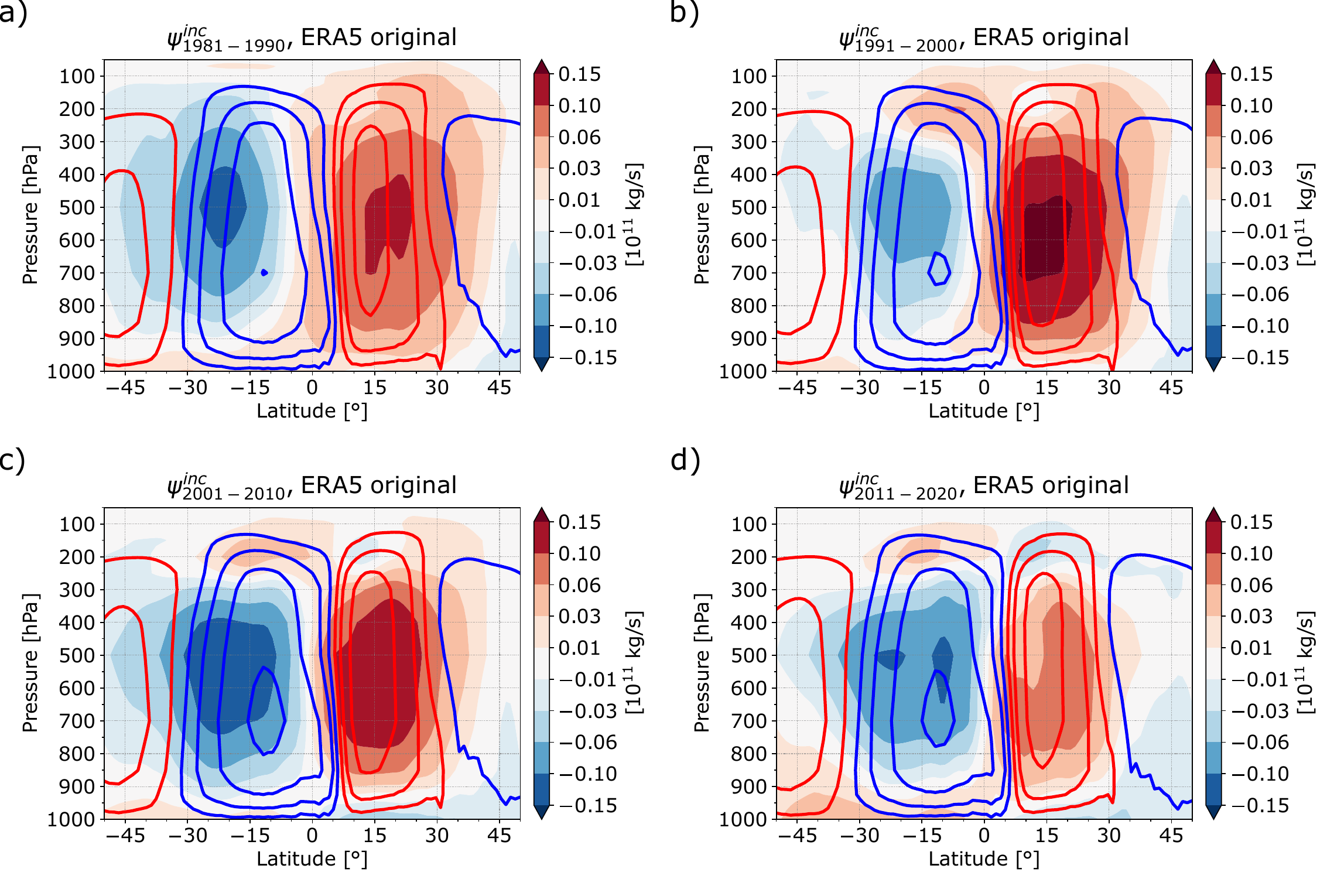}
\caption{Same as Supplementary Information Fig.~\ref{fig:fig_S10}, but computed from raw ERA5 data.}
\label{fig:fig_S11}
\end{figure}
\pagebreak

\clearpage
\begin{figure}[t!]
\centering
\includegraphics[width=0.98\textwidth]{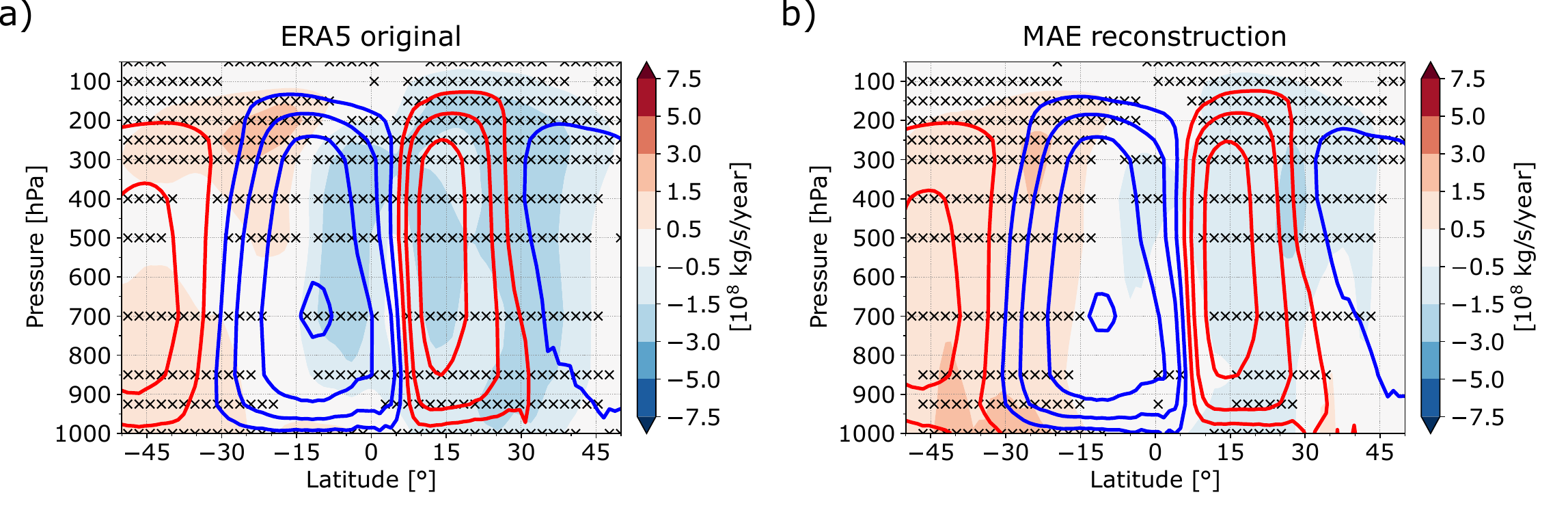}
\caption{\textbf{Comparison of contribution of analysis increments to NHC trends in raw ERA5 reanalysis and MAE reconstructions derived by sampling ERA5 at observation locations.} \textbf{a}, Trends in the annual-mean stream function of analysis increments in raw ERA5 reanalysis data over the 1980–2024 period. \textbf{b}, as in \textbf{a}, but computed from the MAE reconstructions. In both panels, contour lines represent the annual-mean global Hadley circulation, while colours indicate the value of the trend in analysis increments. Red contours denote positive climatological stream function values (0.1, 0.3, 0.6, 1 $\times$ 10$^{11}$ kg s$^{-1}$) and blue contours represent their negative counterparts. Crosses mark regions where trends are statistically significant at the 95\% confidence level, determined using the trend-free pre-whitening Mann-Kendall test.}
\label{fig:fig_S12}
\end{figure}
\pagebreak

\clearpage
\begin{figure}[t!]
\centering
\includegraphics[width=0.98\textwidth]{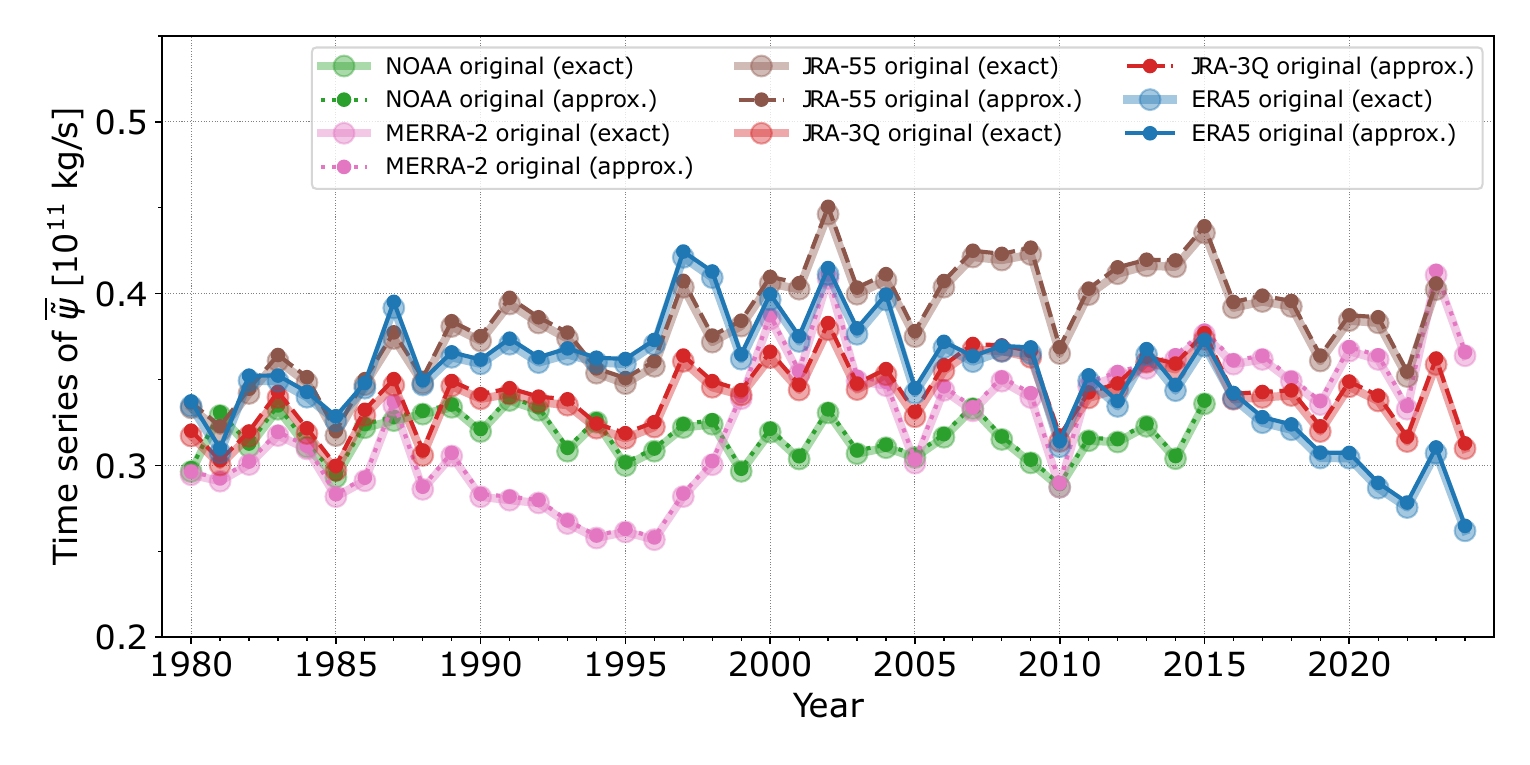}
\caption{\textbf{Validation of the approximate NHC stream-function metric.} Time series of the annual-mean Hadley circulation strength in different reanalyses computed using the meridional average of the stream function, $\langle\psi(\varphi,p)\rangle$, (shading; main text Eq. 9), and its approximation $\tilde{\psi}$(p) (line; main text Eq. 10).}
\label{fig:fig_S13}
\end{figure}
\pagebreak

\clearpage
\begin{figure}[t!]
\centering
\includegraphics[width=0.98\textwidth]{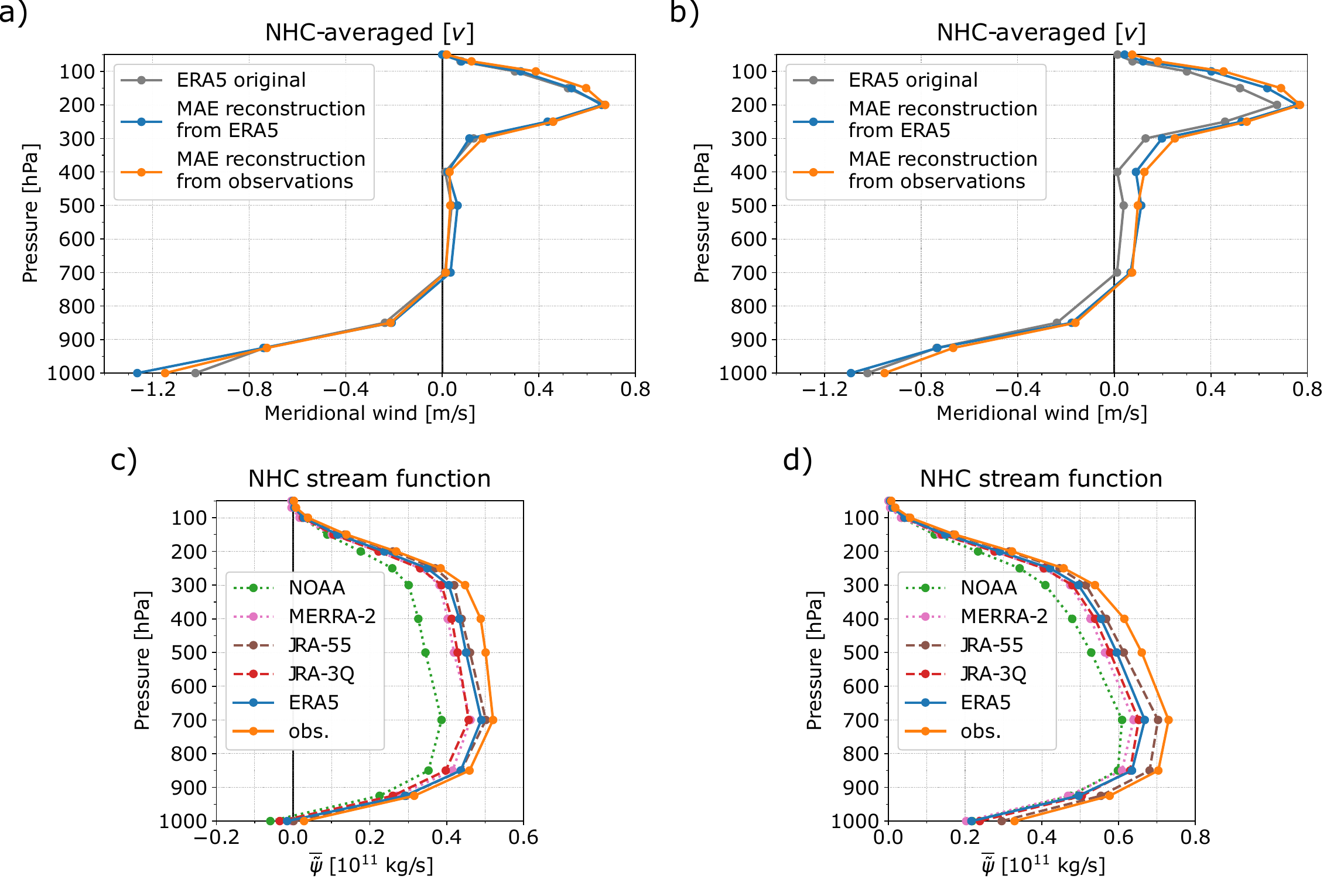}
\caption{\textbf{MAE GNN ensemble member selection based on the physical consistency constraints.} \textbf{a},\textbf{b}, Vertical profile of the zonally averaged meridional wind ($[v]$), averaged over the extent of the NHC (6°N–31°N) over period 1980–2024; \textbf{c},\textbf{d}, Mean stream function profiles representing the climatological NHC strength (ERA5, JRA-3Q, MERRA-2, 1980–2024; JRA-55, 1980–2023; NOAA 20CRv3, 1980–2015). An MAE-GNN version is considered to have passed the selection filter only if its reconstructions consistently satisfy two criteria: low deviation of meridional wind profile relative to the raw target reanalysis (panels \textbf{a},\textbf{b}) and surface mass continuity closure (panels \textbf{c},\textbf{d}), described in Methods (Physical Consistency Constraints). The MAE-GNN version shown on the left (\textbf{a},\textbf{c}) passed these criteria and was retained, while the MAE-GNN version on the right (\textbf{b},\textbf{d}) was discarded.}
\label{fig:fig_S14}
\end{figure}
\pagebreak

\clearpage
\begin{figure}[t!]
\centering
\includegraphics[width=0.7\textwidth]{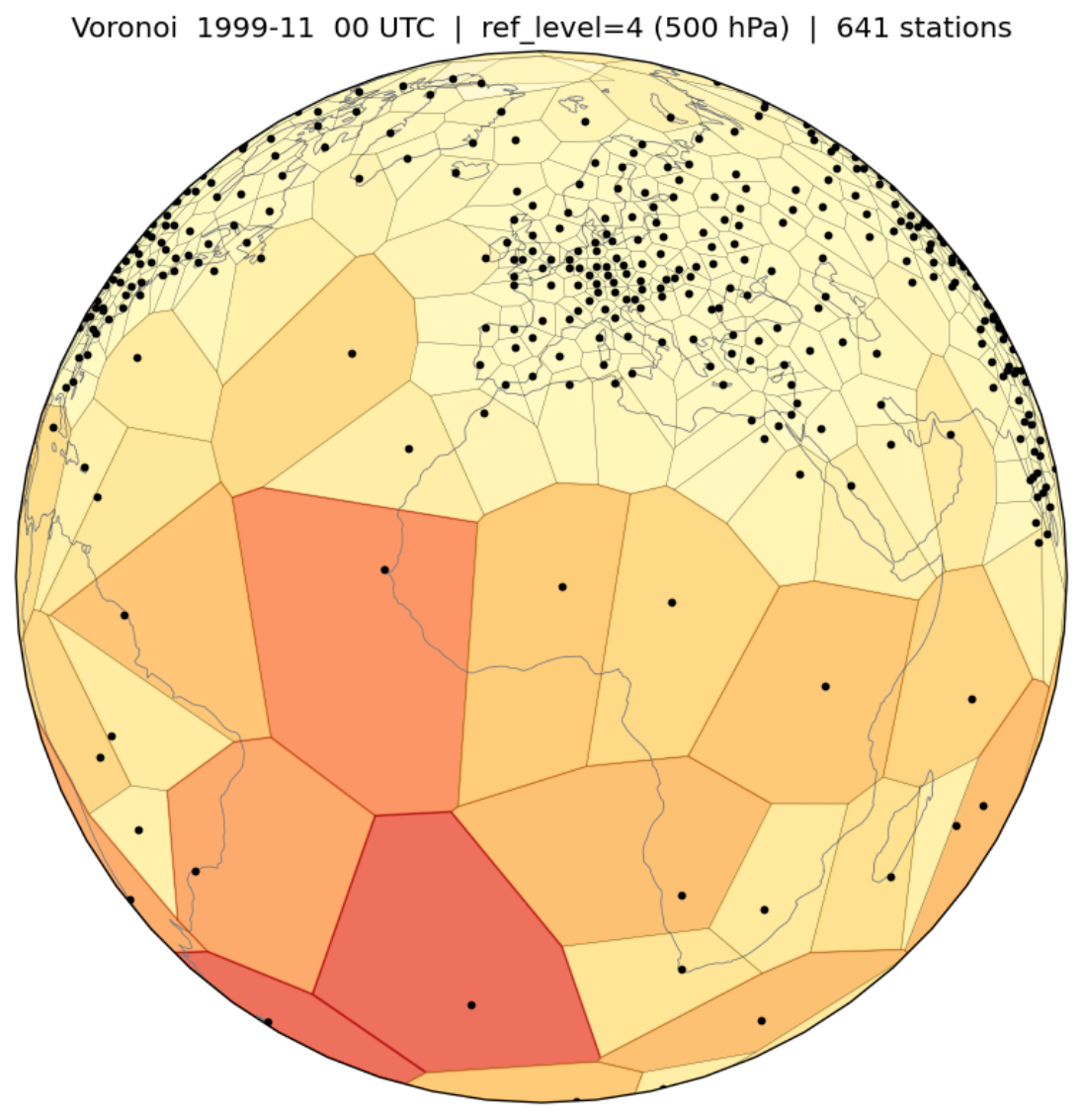}
\caption{\textbf{Spherical Voronoi tessellation of the radiosonde network.} The tessellation is shown for November 1999 at 500 hPa, based on 641 stations used as Voronoi nodes. Black dots indicate station locations, and coloured polygons represent Voronoi cells. Yellow tones correspond to small cells and dense observational coverage, whereas red tones indicate large cells and sparse coverage.}
\label{fig:fig_S15}
\end{figure}
\pagebreak

\clearpage
\begin{figure}[t!]
\centering
\includegraphics[width=0.98\textwidth]{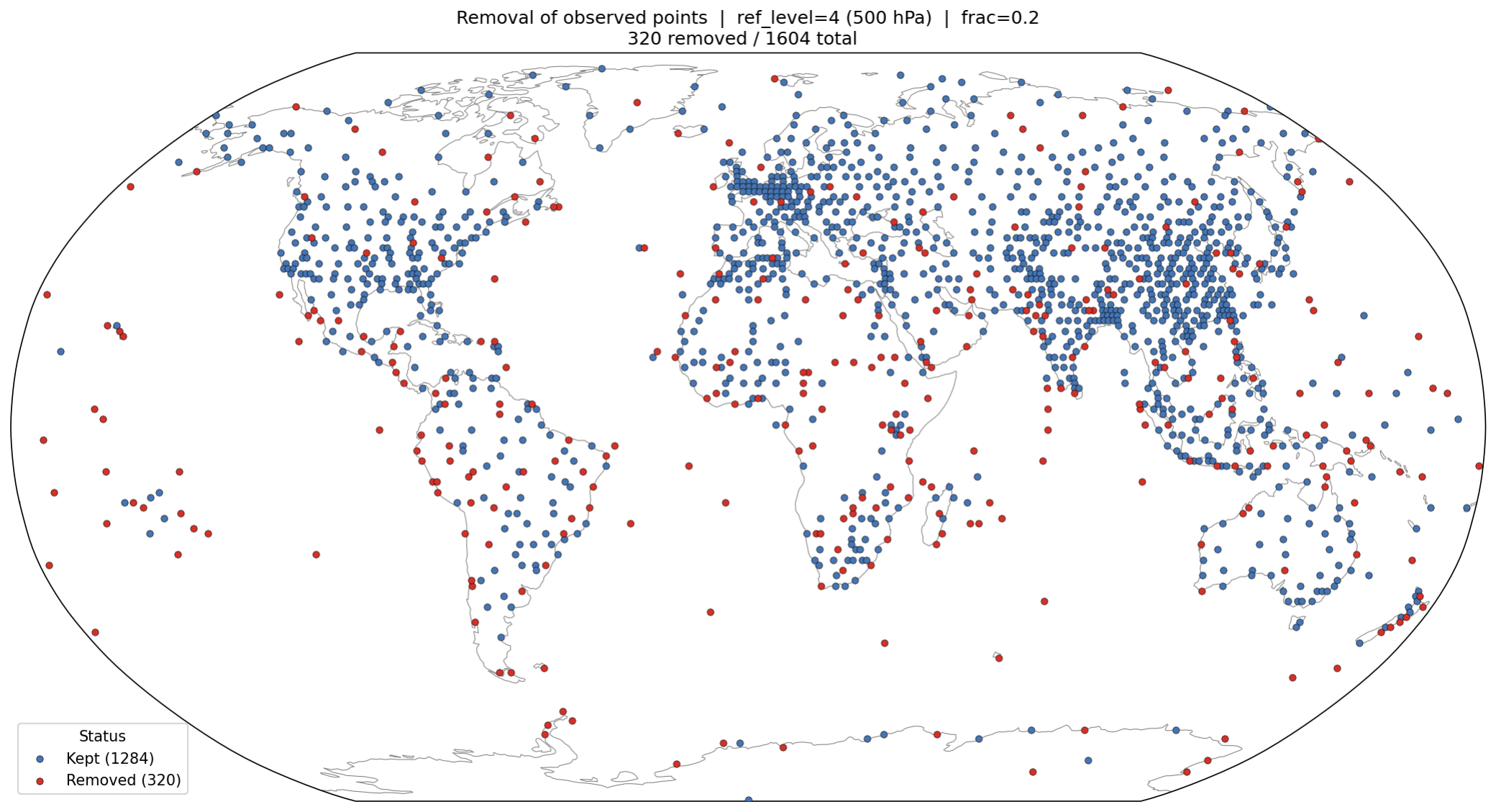}
\caption{\textbf{Removal of observed nodes.} Example of station removal in the radiosonde network, with 320 nodes (20\%) removed out of 1604 across all pressure levels. Removed nodes are shown in red, and retained nodes in blue.}
\label{fig:fig_S16}
\end{figure}